\documentclass[twocolumn,aps,pra,longbibliography,noeprint,nourl,superscriptaddress,amsmath,amssymb]{revtex4-1}

\usepackage{graphicx}
\usepackage{color}
\usepackage[dvipsnames]{xcolor}
\usepackage{upgreek}
\usepackage{float}
\usepackage{upgreek}
\usepackage{amsmath,amssymb}

\begin{document}

\title{Prospects for nuclear spin hyperpolarisation of molecular samples using nitrogen-vacancy centres in diamond} 


\author{J.-P. Tetienne}
\email{jtetienne@unimelb.edu.au}
\affiliation{School of Physics, University of Melbourne, Parkville, VIC 3010, Australia}
\affiliation{Centre for Quantum Computation and Communication Technology, School of Physics, University of Melbourne, Parkville, VIC 3010, Australia}

\author{L. T. Hall}
\affiliation{School of Physics, University of Melbourne, Parkville, VIC 3010, Australia}

\author{A. J. Healey}
\affiliation{School of Physics, University of Melbourne, Parkville, VIC 3010, Australia}

\author{G. A. L. White}
\affiliation{School of Physics, University of Melbourne, Parkville, VIC 3010, Australia}


\author{M.-A. Sani}
\affiliation{School of Chemistry, Bio21 Institute, University of Melbourne, VIC 3010, Australia}

\author{F. Separovic}
\affiliation{School of Chemistry, Bio21 Institute, University of Melbourne, VIC 3010, Australia}

\author{L. C. L. Hollenberg}
\email{lloydch@unimelb.edu.au}
\affiliation{School of Physics, University of Melbourne, Parkville, VIC 3010, Australia}
\affiliation{Centre for Quantum Computation and Communication Technology, School of Physics, University of Melbourne, Parkville, VIC 3010, Australia}

\begin{abstract}

After initial proof-of-principle demonstrations, optically pumped nitrogen-vacancy (NV) centres in diamond have been proposed as a non-invasive platform to achieve hyperpolarisation of nuclear spins in molecular samples over macroscopic volumes and enhance the sensitivity in nuclear magnetic resonance (NMR) experiments. In this work, we model the process of polarisation of external samples by NV centres and theoretically evaluate their performance in a range of scenarios. We find that average nuclear spin polarisations exceeding 10\% can in principle be generated over macroscopic sample volumes ($\gtrsim\mu$L) with a careful engineering of the system's geometry to maximise the diamond-sample contact area. The fabrication requirements and other practical challenges are discussed. 
We then explore the possibility of exploiting local polarisation enhancements in nano/micro-NMR experiments based on NV centres. For micro-NMR, we find that modest signal enhancements over thermal polarisation (by 1-2 orders of magnitude) can in essence be achieved with existing technology, with larger enhancements achievable via micro-structuring of the sample/substrate interface. However, there is generally no benefit for nano-NMR where the detection of statistical polarisation provides the largest signal-to-noise ratio. This work will guide future experimental efforts to integrate NV-based hyperpolarisation to NMR systems.


\end{abstract}

\maketitle 

\section{Introduction}

Nuclear magnetic resonance (NMR) is a powerful technique that can provide structural and dynamic information on molecular objects (NMR spectroscopy) and spatial information with submillimetre resolution for medical diagnosis and in materials science (NMR imaging, or MRI). The NMR signal generally originates from the weak thermal polarisation ($P_{\rm th}$) of nuclear spins at room temperature, for instance $P_{\rm th}\approx10^{-5}$ for protons in a magnetic field of 3 T. This constitutes a major limiting factor to the sensitivity of NMR spectroscopy~\cite{Lee2014}, and consequently to the spatial resolution of MRI. A range of methods has been developed to enhance this polarisation, denoted hyperpolarisation methods, e.g. optical pumping~\cite{Walker1997,Navon1996}, para-hydrogen-induced polarisation (PHIP)~\cite{Natterer1997,Hovener2018}, and dynamic nuclear polarisation (DNP)~\cite{Abragam1978,Wind1985,Hall1997,Tateishi2014,Sani2019}. In essence, all these methods rely on the creation of spin order in a given medium (a noble gas for optical pumping, a dihydrogen gas for PHIP, a solution containing unpaired electrons for DNP) and the subsequent transfer of this spin order to the nuclear spins of the target object. Using these techniques, polarisations ($P_{\rm HP}$) far exceeding $P_{\rm th}$ are routinely achieved, typically from $P_{\rm HP}\approx5\%$ to 90\%. However, they remain technically challenging to apply especially when the target molecules are in solution. For instance, PHIP involves the insertion of para-hydrogen into the target via catalytic hydrogenation, whereas DNP often relies on a freeze-thaw cycle where the polarisation step takes place at cryogenic temperatures (typically below 20~K, or about 100~K for biological NMR applications~\cite{Sani2019}). Consequently, hyperpolarisation techniques remain reserved for specialised applications.  

Recently, the advent of solid-state spin systems originally developed for quantum information science has opened the prospect of a new avenue to achieve hyperpolarisation. In particular, the nitrogen-vacancy (NV) centre in diamond, whose electron spin can be rapidly ($\sim\mu$s) and efficiently polarised ($\approx80\%$) at room temperature by optical pumping, has been proposed as an alternative source of spin order that could be transferred to the target molecules in a similar fashion to DNP~\cite{Broadway2018a,Fernandez2018,Shagieva2018}. Unlike conventional DNP, however, NV centres do not require cryogenic cooling or high magnetic fields, which could potentially enormously simplify the associated infrastructure. Moreover, diamond, being a chemically inert solid, can be relatively easily interfaced with the target molecules with minimum impact. But this also comes with a significant challenge: because the NV centres are relatively sparse and confined in a solid matrix (diamond) that is physically distinct from the target object, the contact area between the polarisation source and the target is drastically reduced compared to standard hyperpolarisation methods which involve a full mixing on molecular scales. 

Proof-of-principle demonstrations of NV-based hyperpolarisation were initially carried out on nuclear spins intrinsic to the diamond~\cite{Henstra2008,London2013,King2010,Alvarez2015,King2015,Pagliero2018,Ajoy2018,Henshaw2019,Schwartz2018,Hovav2018,Lang2019}, and recently hyperpolarisation of molecular spins external to the diamond was demonstrated with a single NV centre via lab-frame cross-relaxation (CR)~\cite{Broadway2018a} and nuclear spin orientation via electron spin locking (NOVEL)~\cite{Fernandez2018,Shagieva2018}. While the polarisation in the immediate vicinity of the NV centre can be quite high ($\approx80\%$) and scaling-up from these initial results appears encouraging~\cite{Broadway2018a}, it remains unclear whether NV-based hyperpolarisation is a viable approach to enhance NMR signals  of macroscopic sample volumes.

In this work, we address this question by modelling the general process of polarisation of an ensemble of nuclear spins in contact with a diamond containing an ensemble of NV centres. We explore a range of scenarios and parameters and find that, even assuming an optimally efficient polarisation transfer, obtaining large enhancements of the average nuclear polarisation requires a careful structuring of the diamond in order to maximise the surface area in contact with the sample, which involves high-aspect-ratio micro-structuring. Practical limitations such as finite NV spin coherence times or finite spin initialisation fidelity, which place further constraints on this requirement, are discussed. We note that the theoretical framework we develop is general and in principle applicable to other solid-state systems hosting electron spins that can be polarised on demand. 

Motivated by the fact that the polarisation may be locally much higher than the average polarisation, we then explore the possibility of enhancing the signal-to-noise ratio (SNR) in both NV-based micro-NMR and nano-NMR experiments, corresponding to sensing volumes of the order of $(10~\mu$m$)^3$ and $(10$~nm$)^3$, respectively~\cite{Glenn2018,Smits2019,Staudacher2013,Mamin2013}. For micro-NMR, we find that modest enhancements over thermal polarisation can be achieved under realistic assumptions, with current technology. For nano-NMR experiments, however, there is generally no SNR enhancement when compared with the standard detection of statistical polarisation. Overall, these results highlight the limitations and challenges of NV-based hyperpolarisation, and as such form the basis for developing a roadmap for future experimental efforts. 

The paper is organised as follows. In Sec.~\ref{sec:singleNV}, we present a theoretical framework to describe the polarisation of an ensemble of nuclear spins using a single NV centre. In Sec.~\ref{sec:convNMR}, this framework is extended to the case of multiple NV centres arranged according to different geometries, and the resulting average polarisation is compared to thermal polarisation. This allows us to determine the influence of the different parameters and identify the requirements to achieve a significant polarisation enhancement. In Sec.~\ref{sec:NV-NMR}, we analyse the case of NV-based micro- and nano-NMR. Finally, in Sec.~\ref{sec:conclusion} we conclude on the prospects on NV-based hyperpolarisation and the experimental challenges ahead.

\section{NV hyperpolarisation of external targets} \label{sec:singleNV}

In this paper, we are interested in the problem of transferring spin polarisation from sparse NV centres in a diamond structure to a comparatively dense ensemble of nuclear spins located outside the diamond (Fig.~\ref{Fig_problem}). This is a quite different situation to that encountered in conventional DNP experiments, and so we will start by discussing the key conceptual challenges for designing a ``diamond hyperpolariser'' (Sec.~\ref{sec:challenges}). We will then describe the dynamics of polarisation exchange between a single NV centre and an ensemble of nuclear spins for a class of optimally efficient protocols (Sec.~\ref{sec:exchange}), and model the resulting build-up of polarisation (Sec.~\ref{sec:buildup}). This model will be extended to the case of multiple NV centres arranged in certain specific geometries in Sec.~\ref{sec:convNMR}. We stress that although we consider the NV centre as the source of polarisation throughout this work, the results are quite general and could apply to any electron spin that can be initialised on demand, with only small corrections needed if the spin quantum number differs from the spin-1 NV system.   


\subsection{Conceptual challenges} \label{sec:challenges}

\begin{figure}[b]
	\includegraphics[width=0.49\textwidth]{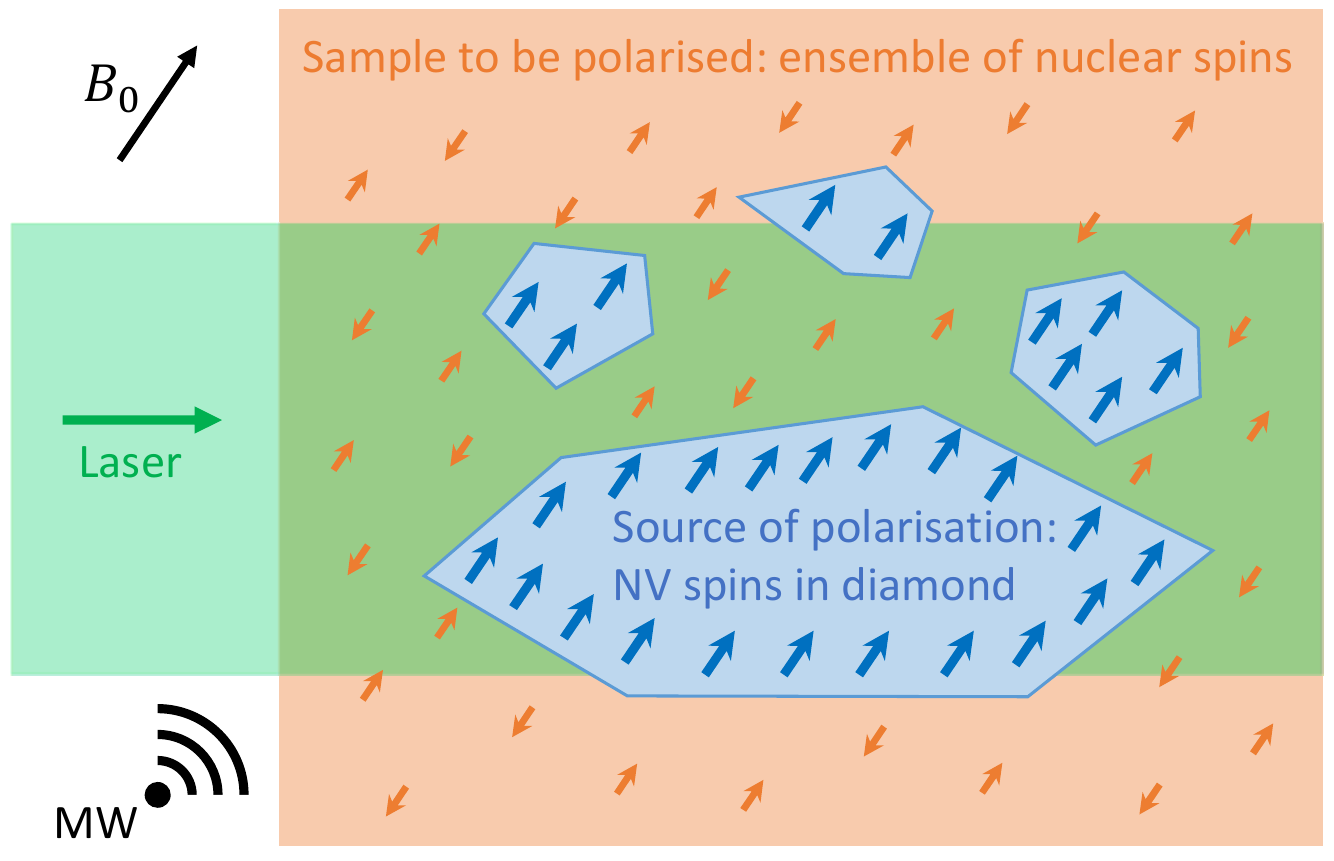}
	\caption{Schematic illustrating the concept of NV-based hyperpolarisation: diamond crystals (blue regions) containing NV spins (blue arrows) are in contact with a sample (orange region) containing nuclear spins (orange arrows). The NV spins are polarised by laser excitation and this polarisation is transferred to the nuclear spins by microwave (MW) irradiation. The quantisation axis is set by the external magnetic field, of strength $B_0$.  
	}  
	\label{Fig_problem}
\end{figure}

The nature of the problem is illustrated in Fig.~\ref{Fig_problem}. Diamond crystals containing NV centres are brought in contact with a sample containing nuclear spins. The goal is to transfer the polarisation from the optically pumped NV spins to the external nuclear spins. Compared to conventional DNP where the polarising agents and the nuclear spins are mixed on the molecular scale, NV-based hyperpolarisation faces two main hurdles. 

The first hurdle is the physical separation between the sample and the NV spins, since the latter are confined inside diamond crystals. This means that there is a minimum distance of several nanometres between the sample and the NV spins closest to the diamond surface, and much more for deeper NVs, when conventional DNP typically involves separations of order $\sim1$~nm or less, consistent throughout the mixture. While the relative number of near-surface NVs can be maximised by choosing an appropriate geometry for the diamond structure, such as a stack of thin diamond plates or an assembly of spherical diamond nanoparticles (see Sec.~\ref{sec:convNMR}), the relatively large minimum gap remains a challenge. Nevertheless, because of their relative isolation the NV spins can have a relatively long lifetime ($T_{1,\rm NV}\approx5$~ms at room temperature~\cite{Jarmola2012}) and coherence time (approaching $T_{1,\rm NV}$ with dynamical decoupling~\cite{Bar-Gill2013}), and as a result polarisation rates comparable to (or exceeding) those of conventional DNP may be achievable.

The second hurdle is that NV centres are typically sparse within a diamond crystal even with aggressive doping techniques ($\sim10$~ppm at most, i.e. roughly one NV per (10~nm)$^3$ volume, see discussion in Sec.~\ref{sec:density}). This means there are very few NV centres available for each nuclear spin, especially if the diamond material occupies a small fraction of the total volume (diamond + sample), as discussed previously in Ref.~\cite{Broadway2018a}. For instance, for a sample containing 1 nuclear spin per (1~nm)$^3$ volume (i.e. a concentration of 1.7~M), there is a maximum of 1 NV per $10^3$ nuclear spins assuming a 1:1 diamond/sample volume ratio (such as in a structure alternating diamond slabs and sample slabs of identical thickness), and 1 NV per $\sim10^6$ nuclear spins if one uses diamond nanoparticles at a concentration of $\sim1$~g/L. In contrast, typical mixtures for DNP have 1 electron spin for just a few nuclear spins to be polarised. 

Thus, NV hyperpolarisation comes with a major disadvantage compared to conventional DNP, caused by the relatively low density and remote nature of the polarising agents. The purpose of this work is to determine whether an efficient polarisation transfer may be able to compensate for this deficit and make NV hyperpolarisation competitive. To this end, we will initially consider an idealised scenario where the NV-nuclear system is perfectly coherent and polarisation can be transferred at a rate simply limited by the strength of the magnetic dipole coupling. This is in contrast with conventional DNP where the mechanisms relied upon (e.g., solid effect, thermal mixing, Overhauser effect, etc.~\cite{Abragam1978,Wind1985}) are designed to operate in a regime dominated by spin relaxation processes leading to transfer rates far below the dipolar coupling limit. Our aim is to use this best-case scenario to assess the potential of NV hyperpolarisation and identify the main challenges and limiting factors. The effect of spin relaxation processes and other practical considerations will be discussed in Sec.~\ref{sec:limitations}.

\subsection{Dynamics of polarisation exchange} \label{sec:exchange}

In this work, we focus on a class of protocols that enable, in the absence of relaxation processes, coherent polarisation transfer (flip-flop) between the NV spin and the target nuclear spin ensemble. This class includes CR~\cite{Broadway2018a} and NOVEL~\cite{Henstra2008,London2013,Fernandez2018,Shagieva2018}, which rely on a continuous interaction, as well as pulsed protocols such as PulsePol~\cite{Schwartz2018}, refocused NOVEL~\cite{Hovav2018} and PolCPMG~\cite{Lang2019}. With the exception of CR which is microwave-free, all these protocols require microwave excitation resonant with the NV electron spin transition, and the flip-flop interaction is enabled by careful tuning of the microwave amplitude and/or pulsing parameters. Other protocols relying on thermal mixing, the cross effect or the solid effect were demonstrated~\cite{King2010,Alvarez2015,King2015,Pagliero2018,Ajoy2018,Henshaw2019}, but they are designed to operate in a regime dominated by relaxation processes and thus are not considered here.

\begin{figure}[b]
	\includegraphics[width=0.49\textwidth]{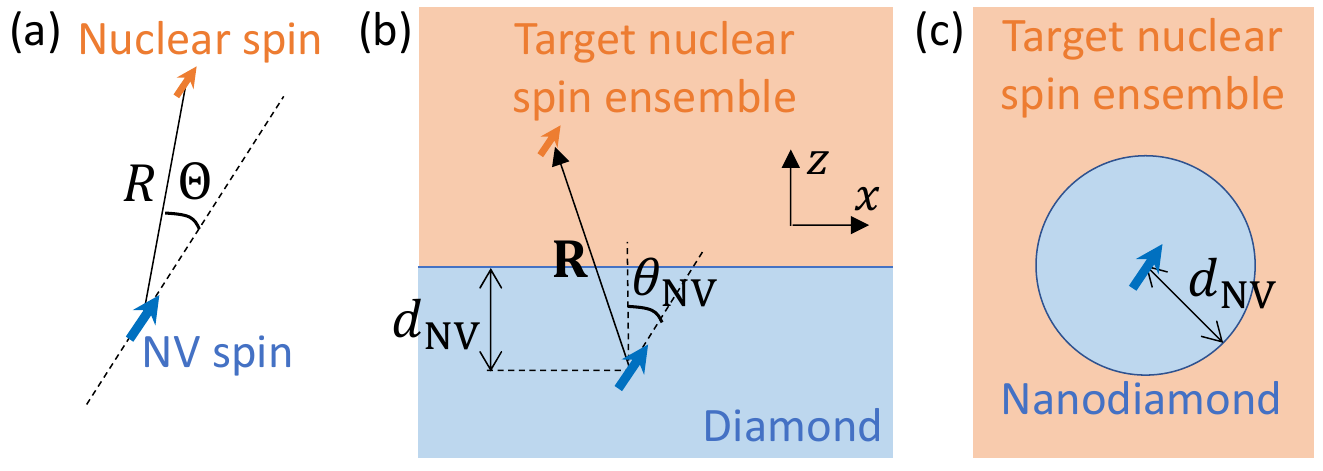}
	\caption{Definitions of the geometrical parameters in the case of (a) a single NV spin (blue arrow) interacting with a single nuclear spin (orange arrow), (b) a single NV spin interacting with a semi-infinite slab of nuclear spins (orange box), and (c) a single NV spin at the centre of a spherical nanodiamond (blue sphere) immersed in an infinite medium of nuclear spins. In (b), the orange arrow represents one spin among the ensemble, to define the position vector {\bf R}.
	}  
	\label{Fig_definitions}
\end{figure}

Let us consider the case of a single NV centre interacting with a single nuclear spin located at a distance $R$, with a polar angle $\Theta$ relative to the quantisation axis (Fig.~\ref{Fig_definitions}a). We assume that the NV centre's electron spin (with gyromagnetic ratio $\gamma_e$) is perfectly initialised in the $|0\rangle$ state. The nuclear spin is a spin-$\frac{1}{2}$ with gyromagnetic ratio $\gamma_n$. We denote $p_\uparrow$ ($p_\downarrow$) as the probability for the nuclear spin to be in the $|\uparrow\rangle$ ($|\downarrow\rangle$) state. The spin polarisation is then defined as $P=p_\uparrow-p_\downarrow$. 

For the class of coherent protocols mentioned above, the polarisation $P$ evolves as~\cite{Broadway2018a,Hall2020}
\begin{eqnarray}
P(\tau)=P(0)+[1-P(0)]\sin^2\left(\frac{A_{\rm s}\tau}{2}\right)
\end{eqnarray}  
where $\tau$ is the interaction time and $A_{\rm s}$ is the protocol-dependent flip-flop rate. Generally, CR offers the largest flip-flop rate followed by NOVEL and PulsePol, as shown in Ref.~\cite{Hall2020}. However, CR and NOVEL are not expected to operate in the coherent regime for external samples due to these protocols' sensitivity to surface-induced NV spin dephasing. Therefore, for all numerical evaluations, we will consider the PulsePol protocol as it is the most robust against dephasing and other imperfections~\cite{Schwartz2018} hence the best candidate to realise this idealised coherent scenario. Namely, for PulsePol the flip-flop rate is~\cite{Schwartz2018,Hall2020}
\begin{eqnarray} \label{eq:Asingle}
A_{\rm s}=\frac{3\alpha a}{2R^3}|\sin(\Theta)\cos(\Theta)|
\end{eqnarray}  
where $a = \frac{\mu_0\hbar\gamma_e\gamma_n}{4\pi}$ is the magnetic dipole coupling constant and $\alpha\approx0.72$ is a numerical factor. 

At time $\tau=\tau_0\equiv\frac{\pi}{A_{\rm s}}$, which we will refer to as the flip-flop time, the nuclear spin is fully polarised ($P=1$) corresponding to a pure spin state $|\uparrow\rangle$. Thus, it takes a duration $\tau_0$ to transfer a fraction (dependent on $P(0)$) of a single quantum of angular momentum from the NV spin to the nuclear spin. As an example, in this idealised scenario it would take $\tau_0\approx1.6$~ms to fully polarise a single proton spin ($^1$H) at a distance $R=5$~nm and angle $\Theta=54.7^\circ$, using PulsePol. 



If the NV interacts with an ensemble of independent nuclear spins with uniform density $\rho_n$, it will still exchange polarisation coherently with the ensemble but at a faster rate $A_0$ given by~\cite{Broadway2018a}
\begin{eqnarray} \label{eq:Atot}
A_0^2 & = & \rho_n\int A_{\rm s}^2({\bf R}){\rm d}^3{\bf R}
\end{eqnarray} 
where $A_{\rm s}({\bf R})$ is the coupling strength for a spin located at position {\bf R}, which is given by Eq.~\ref{eq:Asingle} in the case of PulsePol. As before, the flip-flop time is defined as $\tau_0\equiv\frac{\pi}{A_0}$. 

The rate $A_0$ can be evaluated analytically for certain geometries. Let us first consider the case of a semi-infinite slab of nuclear spins placed on a flat diamond surface (Fig.~\ref{Fig_definitions}b). The NV centre is located at a depth $d_{\rm NV}$ below the diamond surface and its spin quantisation axis forms an angle $\theta_{\rm NV}$ with the surface normal (defined as the $z$ axis). In this case, we obtain
\begin{eqnarray} \label{eq:Atot_flat}
A_0^2 & = & \frac{\rho_n\alpha^2 a^2 \pi [55+12\cos(2\theta_{\rm NV})-3\cos(4\theta_{\rm NV})] }{1024 d_{\rm NV}^3}~. 
\end{eqnarray} 
As an example, for $^1$H spins in frozen water ($\rho_n=66$~nm$^{-3}$) with $d_{\rm NV}=5$~nm and $\theta_{\rm NV}=54.7^\circ$, we get a flip-flop time $\tau_0\approx30~\mu$s.

Alternatively, consider an NV centre located at the centre of a spherical diamond nanoparticle (nanodiamond) immersed in the nuclear spin ensemble (Fig.~\ref{Fig_definitions}c). If $d_{\rm NV}$ is the radius of the nanodiamond, Eq.~\ref{eq:Atot} evaluates to
\begin{eqnarray} \label{eq:Atot_sphere}
A_0^2 & = & \frac{2\pi\rho_n\alpha^2 a^2}{5 d_{\rm NV}^3}~.
\end{eqnarray} 
For $^1$H in frozen water with $d_{\rm NV}=5$~nm, we obtain $\tau_0\approx11~\mu$s, which is nearly 3 times shorter than with the flat surface.

From Eq.~\ref{eq:Atot_flat} and \ref{eq:Atot_sphere}, it follows that the flip-flop time scales as $\tau_0=\frac{\pi}{A_0}\propto d_{\rm NV}^{3/2}\rho_n^{-1/2}\gamma_n^{-1}$. For the flat surface geometry, a shallower NV will give a shorter flip-flop time, e.g. $\tau_0\approx8~\mu$s for a 2-nm deep NV, which is often considered a practical lower limit~\cite{Kaviani2014,Loretz2014}. Meanwhile, more dilute samples will lead to longer times, e.g. $\tau_0\approx310~\mu$s for 1~M of $^1$H spins in a deuterated solvent ($\rho_n=0.6$~nm$^{-3}$, $d_{\rm NV}=5$~nm), and 1~M of $^{13}$C spins would give $\tau_0\approx1.2$~ms. These numbers illustrate the typical time scales involved in NV-based hyperpolarisation. 


It is important to note that the flip-flop time $\tau_0$ corresponds to the transfer of a single (at most) quantum of angular momentum, provided by the NV spin. For an ensemble of nuclear spins, this transferred momentum must thus be shared among the nuclear spins, proportionally to their respective coupling strength. Precisely, the polarisation of a nuclear spin at position {\bf R}, under a continuum approximation~\cite{Broadway2018a}, evolves as
\begin{eqnarray} \label{eq:PRtau}
P({\bf R},\tau)=P({\bf R},0)+[1-P({\bf R},0)]\frac{A_{\rm s}^2({\bf R})}{A_0^2}\sin^2\left(\frac{A_0\tau}{2}\right). \nonumber  \\
\end{eqnarray}  
To build up polarisation, the NV therefore needs to be reinitialised and the protocol repeated, as we analyse below.

\subsection{Modelling the polarisation build-up} \label{sec:buildup}

We consider the situation where the polarisation transfer protocol is continuously repeated. That is, we continuously repeat a cycle consisting of initialising the NV spin to $|0\rangle$ and applying the protocol for a duration $\tau$. Each cycle increases the polarisation by a small amount, as described by Eq.~\ref{eq:PRtau}. The polarisation after each cycle of duration $\tau$, which we denote as $P({\bf R},t)$ where $t$ is the total time (a multiple of $\tau$), can then be described by a differential equation $\frac{\partial P}{\partial t}=u(1-P)$ where $u$ is the position-dependent polarisation rate, or ``cooling'' rate, given by 
\begin{eqnarray} \label{eq:uR}
u({\bf R}) & = & \frac{P({\bf R},t+\tau)-P({\bf R},t)}{\tau[1-P({\bf R},t)]} \nonumber \\
& = & \frac{1}{\tau}\frac{A_{\rm s}^2({\bf R})}{A_0^2}\sin^2\left(\frac{A_0\tau}{2}\right).
\end{eqnarray} 
The optimum duration $\tau$ to maximise $u({\bf R})$ is $\tau_{\rm opt}\approx0.74\tau_0$ and gives $u({\bf R})\approx\frac{1.14}{\tau_0}\frac{A_{\rm s}^2({\bf R})}{A_0^2}$. The differential equation produced can be extended to include spin-lattice relaxation of the nuclear spins (relaxation rate $\Gamma_{1,n}$) as well as polarisation diffusion (diffusion coefficient $D_n$), leading to the differential equation~\cite{Broadway2018a}
\begin{widetext}
\begin{eqnarray} \label{eq:master}
\frac{\partial P({\bf R},t)}{\partial t}=u({\bf R})[1-P({\bf R},t)]-\Gamma_{1,n}P({\bf R},t)+D_n\nabla^2 P({\bf R},t)~.
\end{eqnarray} 
\end{widetext}
The diffusion term may capture molecular diffusion in the case of a liquid sample (assuming diffusion is slow enough that the flip-flop dynamics described in the previous section remains approximately valid, see Sec.~\ref{sec:diffusion}), or dipole-mediated spin diffusion in the case of a solid sample.  

\begin{figure}[b]
	\includegraphics[width=0.45\textwidth]{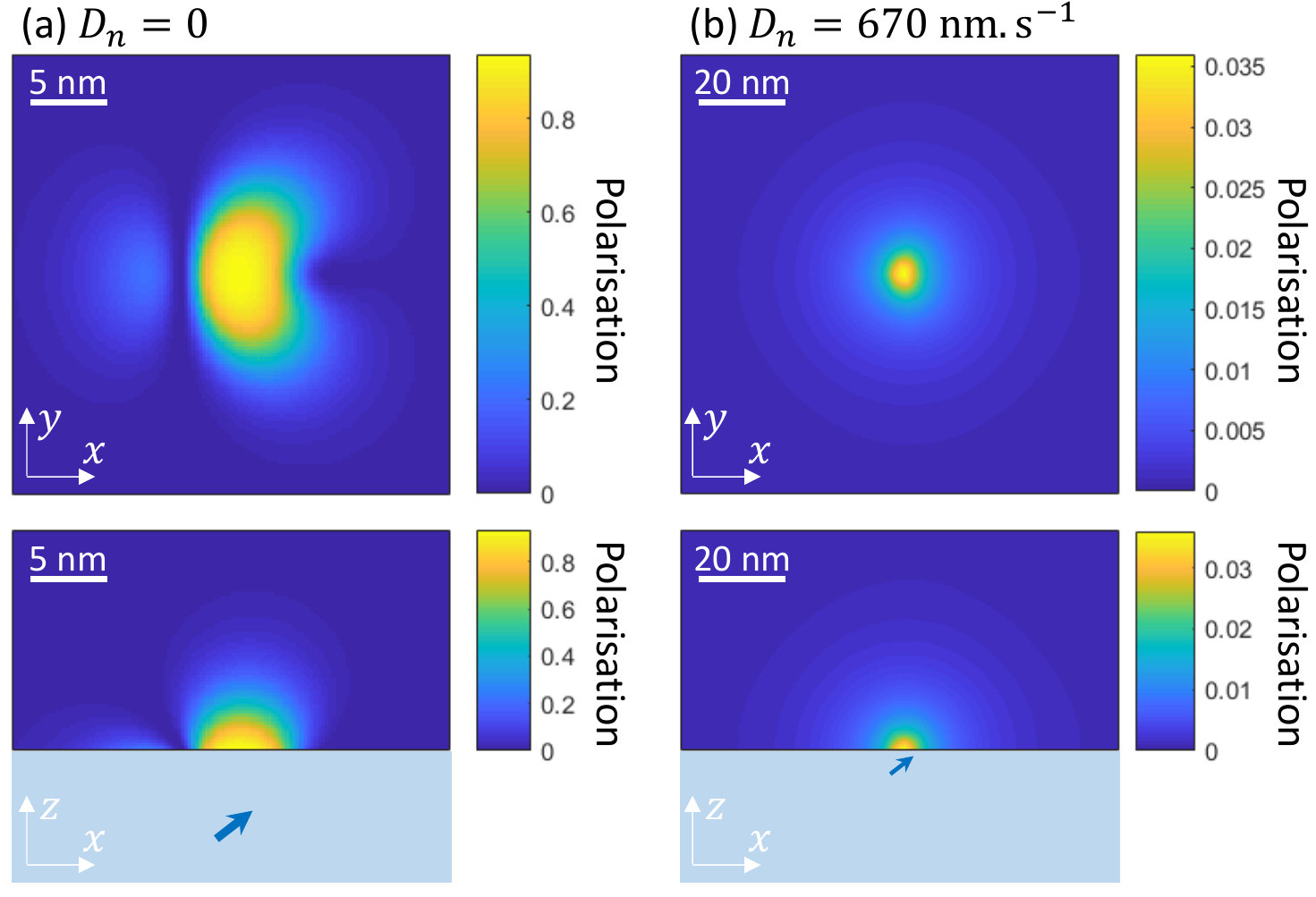}
	\caption{Calculated polarisation maps in the steady state ($P({\bf R},\infty)$) in the $xy$ plane at the diamond-sample interface (top) and in the $xz$ plane encompassing the NV (bottom) without diffusion (a) and with $D_n=670$~nm$^2$s$^{-1}$ (b). The NV spin has a depth $d_{\rm NV}=5$~nm and is oriented at an angle $\theta_{\rm NV}=54.7^\circ$ in the $xz$ plane. The nuclear spins ($^1$H) have a density $\rho_n=66$~nm$^{-3}$ and a relaxation time $T_{1,n}=1$~s. The PulsePol protocol is applied with an interaction time $\tau=\tau_{\rm opt}=22~\mu$s.
	}  
	\label{Fig_PolMaps}
\end{figure}

Before solving Eq.~\ref{eq:master} numerically in the general case, it is useful to examine the solution when diffusion is neglected ($D_n=0$),
\begin{eqnarray} \label{eq:Pt}
P({\bf R},t)=P({\bf R},\infty)\left[1-e^{(u({\bf R})+\Gamma_{1,n})t}\right]
\end{eqnarray}
where we assumed that the sample is initially unpolarised, $P({\bf R},0)=0$, and the steady-state value is given by
\begin{eqnarray} \label{eq:Psat}
P({\bf R},\infty)=\frac{u({\bf R})}{u({\bf R})+\Gamma_{1,n}}.
\end{eqnarray}
This steady-state value is reached in a time of the order of the relaxation time $T_{1,n}=1/\Gamma_{1,n}$, or less if $u({\bf R})\gg\Gamma_{1,n}$. As an example, for $^1$H spins in a dense ensemble ($\rho_n=66$~nm$^{-3}$) with $d_{\rm NV}=5$~nm and $\theta_{\rm NV}=54.7^\circ$, $u({\bf R})$ reaches up to $\approx15$~s$^{-1}$ at some positions ${\bf R}$, leading to a maximum steady-state polarisation of $\approx94\%$ if $T_{1,n}=1$~s. Cross-sections of the polarisation distribution for this scenario are shown in Fig.~\ref{Fig_PolMaps}a, revealing a multi-lobe structure originating from the angular dependence of the dipolar interaction (Eq.~\ref{eq:Asingle}). 

Diffusion acts to spread the polarisation further away from the source (i.e. the NV), but the time scale to reach the steady state remains relatively unchanged, constrained by $T_{1,n}$. Using the approximate expression for a cubic lattice, $D_n\approx0.22\frac{\mu_0}{4\pi}\hbar\gamma_n^2\rho_n^{1/3}$~\cite{Cheung1981}, we obtain $D_n\approx670$~nm$^2$s$^{-1}$ under the same assumptions as before. The spatial extent of the polarisation is thus expected to be of the order of $\sqrt{D_n T_{1,n}}\approx30$~nm. Polarisation maps for this scenario are shown in Fig.~\ref{Fig_PolMaps}b where the polarisation indeed extends over 10's of nanometres, much further than without diffusion (Fig.~\ref{Fig_PolMaps}a), while the maximum polarisation is reduced to $<4\%$. 

A figure of merit to quantify the efficiency of the polarisation process is the effective number of polarised spins, defined as
\begin{eqnarray} \label{eq:Neff}
N_{\rm s}=\rho_n\int P({\bf R},\infty){\rm d}^3{\bf R}~.
\end{eqnarray} 
In the conditions of Fig.~\ref{Fig_PolMaps}, we obtain $N_{\rm s}\approx 20,000$ without diffusion, and $N_{\rm s}\approx 37,000$ with $D_n=670$~nm$^2$s$^{-1}$. The difference is due to a saturation effect: as $P$ approaches 1, the rate of change $\frac{\partial P}{\partial t}$ decreases (see Eq.~\ref{eq:master}); therefore, by keeping the local polarisation to a low level, diffusion allows for a higher total polarisation to be reached. 

\begin{figure}[t!]
	\includegraphics[width=0.49\textwidth]{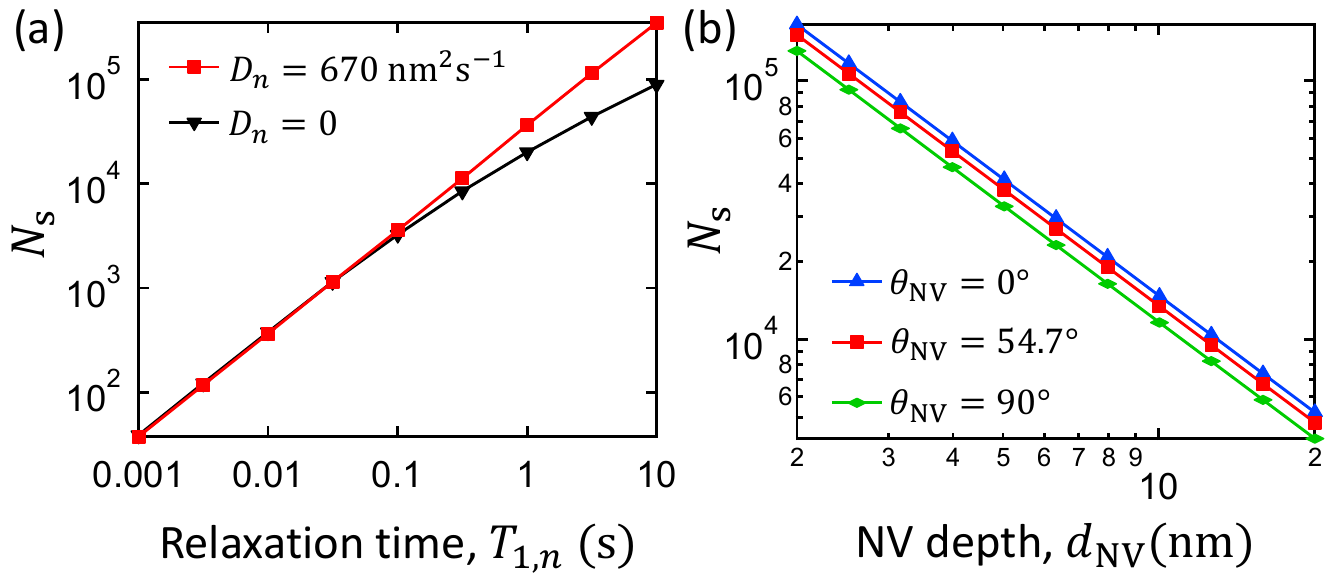}
	\caption{(a) Calculated effective number of polarised spins in the steady state, $N_{\rm s}$, as a function of $T_{1,n}$ for $^1$H spins with $\rho_n=66$~nm$^{-3}$ and $D_n=670$~nm$^2$s$^{-1}$ (red data) or $D_n=0$ (black). The NV spin is such that $d_{\rm NV}=5$~nm and $\theta_{\rm NV}=54.7^\circ$. (b) $N_{\rm s}$ as a function of $d_{\rm NV}$ for $^1$H spins with $\rho_n=66$~nm$^{-3}$, $D_n=670$~nm$^2$s$^{-1}$ and $T_{1,n}=1$~s. The NV angle is $\theta_{\rm NV}=0^\circ$ (blue data), $54.7^\circ$ (red) and $90^\circ$ (green).     
	}  
	\label{Fig_Neff_ideal}
\end{figure}

The effect of diffusion is further illustrated in Fig.~\ref{Fig_Neff_ideal}a which plots $N_{\rm s}$ with and without diffusion, as a function of $T_{1,n}$. For $T_{1,n}\lesssim0.1$~s, diffusion has little effect on $N_{\rm s}$ because the polarisation levels remain $\ll 1$, but for larger $T_{1,n}$ we see that $N_{\rm s}$ becomes comparatively smaller without diffusion due to this saturation effect. Meanwhile, with diffusion $N_{\rm s}$ grows linearly with $T_{1,n}$, as anticipated from Eq.~\ref{eq:Psat} which gives $P({\bf R},\infty)\approx u({\bf R})T_{1,n}$ in the limit $u({\bf R})\ll\Gamma_{1,n}$ (i.e. far from saturation). In this limit, evaluating Eq.~\ref{eq:Neff} using the expression for $u({\bf R})$ from Eq.~\ref{eq:uR} gives
\begin{eqnarray} \label{eq:Neff2}
N_{\rm s}\approx1.14\frac{T_{1,n}}{\tau_0}
\end{eqnarray} 
where we chose the optimal time, $\tau=\tau_{\rm opt}$. Thus, we find, quite as expected, that the number of polarised spins is roughly the maximum build-up time ($\sim T_{1,n}$) divided by the time it takes to polarise one spin ($\tau_0$). Evaluating $\tau_0$ using Eq.~\ref{eq:Atot_flat} or \ref{eq:Atot_sphere}, we obtain  
\begin{eqnarray} \label{eq:Neff_scaling}
N_{\rm s} & \approx & g aT_{1,n}\rho_n^{1/2} d_{\rm NV}^{-3/2} \\
& \propto & \gamma_n T_{1,n}\rho_n^{1/2} d_{\rm NV}^{-3/2}
\end{eqnarray} 
where $g$ is a dimensionless factor of order unity, which depends on the geometry considered (and on the protocol eventually), for instance $g\approx1.14\alpha\left(\frac{2}{5\pi}\right)^{1/2}\approx0.29$ with PulsePol for the nanodiamond geometry (Fig.~\ref{Fig_definitions}c).

For the flat surface geometry (Fig.~\ref{Fig_definitions}b), $g$ depends on the NV angle $\theta_{\rm NV}$. This is illustrated in Fig.~\ref{Fig_Neff_ideal}b which plots $N_{\rm s}$ as a function of the NV depth $d_{\rm NV}$ varying from 2~nm to 20~nm, for different angles $\theta_{\rm NV}$. We see that a maximum polarisation is achieved with $\theta_{\rm NV}=0^\circ$ corresponding to the NV axis along $z$ (blue data), followed by $\theta_{\rm NV}=54.7^\circ$ and $\theta_{\rm NV}=90^\circ$ which are about 10\% and 30\% less efficient, respectively. Nevertheless, these differences are small compared to the effect of $d_{\rm NV}$, as expected from the $N_{\rm s}\propto d_{\rm NV}^{-3/2}$ scaling. In the following we will assume $\theta_{\rm NV}=54.7^\circ$, which is the most commonly found angle as it corresponds to a (100)-oriented diamond surface.

\section{NV-based hyperpolarisation for conventional NMR} \label{sec:convNMR}

In the previous section, we developed a framework to predict the maximum number of spins that can be polarised by a single NV spin, $N_{\rm s}$ (Eq.~\ref{eq:Neff_scaling}). Here, we analyse how macroscopic ensembles of NV centres can be arranged to produce a sizeable polarisation over a sample volume compatible with conventional NMR ($\sim~\mu$L to mL). We analyse two different architectures for such a diamond hyperpolariser, and for each we determine the requirements to achieve a polarisation enhancement.


\subsection{Slab architecture} \label{sec:slab}

\begin{figure}[b]
	\includegraphics[width=0.49\textwidth]{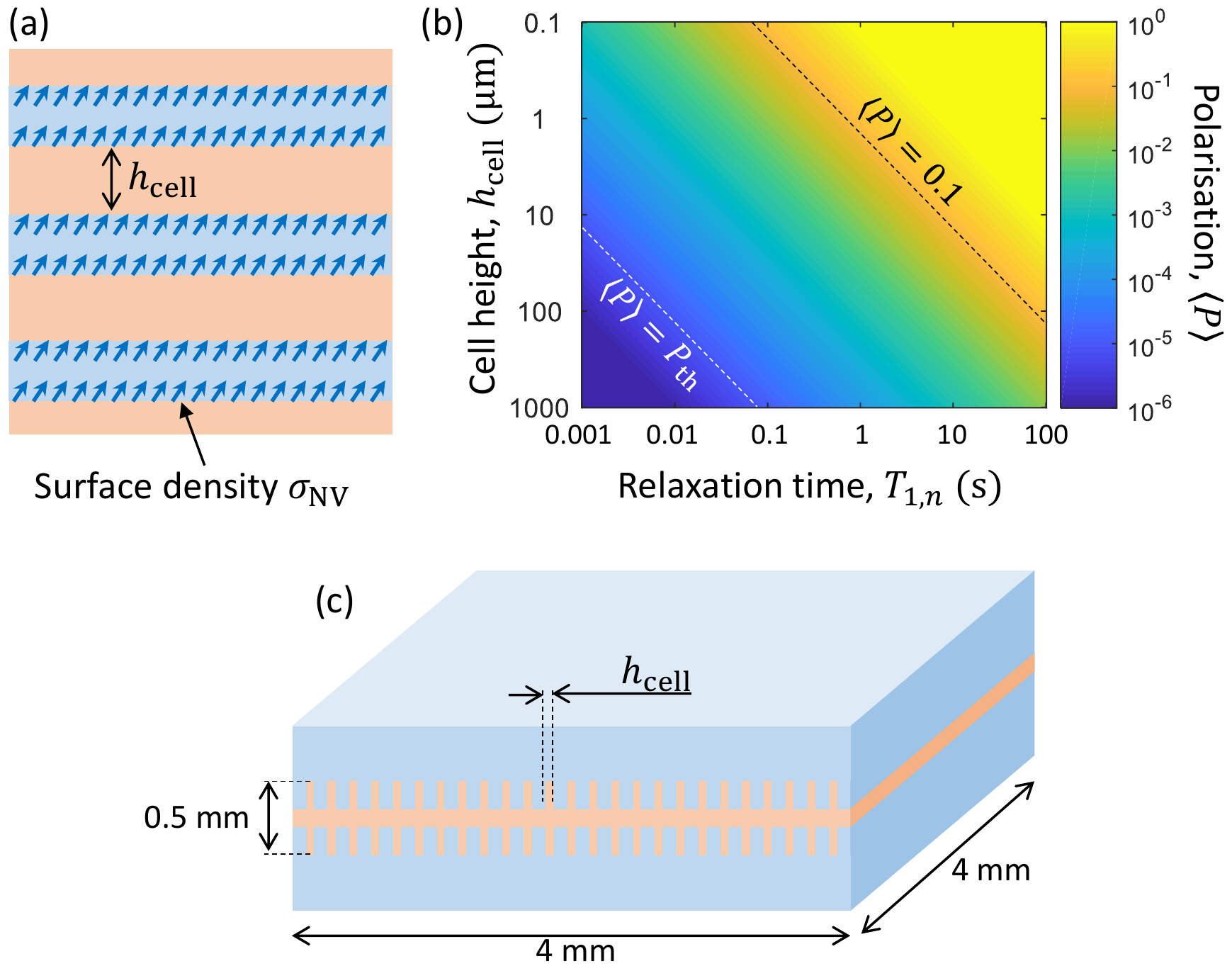}
	\caption{(a) A slab architecture for the hyperpolarisation of a macroscopic sample (orange regions) based on a stack of diamond slabs (blue) with near-surface NV centres (blue arrows) on both sides. (b) Average polarisation induced by the NVs, $\langle P \rangle$, as a function of $T_{1,n}$ and $h_{\rm cell}$, in the geometry (a) for $^1$H spins. The other parameters are: $\sigma_{\rm NV}=10^{16}$~m$^{-2}$, $d_{\rm NV}=5$~nm, $\rho_n=0.6$~nm$^{-3}$. The white (black) dashed line indicates $\langle P \rangle = P_{\rm th}$ ($\langle P \rangle=0.1$). $\langle P \rangle$ is calculated using Eq.~\ref{eq:PNV_slab}, which assumes that the polarisation is $\ll 1$ at any point in the sample. Thus, the plot is not meant to be accurate in the region where $\langle P \rangle$ approaches unity. Where Eq.~\ref{eq:PNV_slab} predicts $\langle P \rangle>1$, the value was capped to 1. (c) Example of a diamond hyperpolariser based on the geometry (a). Two diamond plates are structured to feature $h_{\rm cell}$-wide grooves and sealed together. Example dimensions are indicated, yielding a total sample volume of $\approx 5~\mu$L.   
	}  
	\label{Fig_slab}
\end{figure}

We first consider the geometry proposed in Ref.~\cite{Broadway2018a} and depicted in Fig.~\ref{Fig_slab}a, which employs a stack of diamond slabs comprising arrays of near-surface NV centres on each side, with an areal density $\sigma_{\rm NV}$. The NVs are located at a depth $d_{\rm NV}$ from the diamond surface and form an angle $\theta_{\rm NV}$ with the $z$ axis (normal to the diamond surface). The gap between diamond slabs, filled with the sample to be polarised, is $h_{\rm cell}$. The unit cell of this structure is therefore a slab of sample of thickness $h_{\rm cell}/2$, polarised by a single layer of NV spins. If $N_{\rm s}$ is the number of polarised spins due to a single NV spin within the array, the polarisation will be, when averaged over the entire sample,
\begin{eqnarray} \label{eq:PNV_slab}
\langle P \rangle & = & \frac{2\sigma_{\rm NV}N_{\rm s}}{\rho_n h_{\rm cell}} \\
& = & \frac{2\sigma_{\rm NV}g aT_{1,n}}{\rho_n^{1/2}d_{\rm NV}^{3/2}h_{\rm cell}}.
\end{eqnarray} 
where the second line used Eq.~\ref{eq:Neff_scaling}, $g$ being the geometric factor for the slab geometry ($g\approx0.11$ for $\theta_{\rm NV}=54.7^\circ$).

We compare this NV-induced polarisation with the thermal polarisation~\cite{Lee2014},
\begin{eqnarray} \label{eq:Pth}
P_{\rm th}=\tanh\left(\frac{\hbar\gamma_nB_0}{2k_BT}\right)
\end{eqnarray} 
where $B_0$ is the magnetic field, $T$ the temperature and $k_B$ the Boltzmann constant. In the following, we will compare $\langle P \rangle$ to the thermal polarisation obtained at $B_0=3$~T and $T=300$~K, i.e. $P_{\rm th}\approx10^{-5}$ for $^1$H spins. The 3~T value was chosen to be representative of conventional NMR/MRI experiments, keeping in mind that $P_{\rm th}$ scales approximately as $P_{\rm th}\propto B_0$ so that a 9-T NMR spectrometer, for example, would give a polarisation 3 times as large as that assumed here. We note that $P_{\rm th}$ and $\langle P \rangle$ (via the constant $a$) are both proportional to the gyromagnetic ratio $\gamma_n$, therefore the ratio $\langle P \rangle/P_{\rm th}$ is independent of the nucleus considered.  

Let us examine how the parameters in Eq.~\ref{eq:PNV_slab} can be optimised to achieve $\langle P \rangle \gg P_{\rm th}$. The NV density $\sigma_{\rm NV}$ must be maximised, but is limited by materials considerations to maximum values of the order of $\sigma_{\rm NV}\sim10^{16}$~m$^{-2}$ (see discussion in Sec.~\ref{sec:density}). The NV depth $d_{\rm NV}$ must be minimised but is also limited to a minimum of a few nanometres typically. The density $\rho_n$ depends on the sample to be polarised but has a relatively weak effect ($\rho_n^{1/2}$ scaling) compared to other parameters. The remaining parameters are $T_{1,n}$, which can vary over several orders of magnitude depending on the sample, and $h_{\rm cell}$ which depends on the engineering of the system and could vary from mm to $\mu$m scales. To explore the parameter space, we therefore vary $T_{1,n}$ and $h_{\rm cell}$ while fixing the other parameters to nominal typical values: $\sigma_{\rm NV}=10^{16}$~m$^{-2}$, $d_{\rm NV}=5$~nm, $\rho_n=0.6$~nm$^{-3}$ (i.e. 1~M). 

The outcome of this parameter sweep is presented in Fig.~\ref{Fig_slab}b. The white dashed line corresponds to the $\langle P \rangle = P_{\rm th}$ condition, implying that there is no polarisation enhancement below this line, whereas the black dashed line indicates a large enhancement, $\langle P \rangle = 0.1\approx 10^4 P_{\rm th}$. With $h_{\rm cell}=100~\mu$m, we obtain a modest enhancement $\langle P \rangle \approx 100 P_{\rm th}$ with $T_{1,n}=1$~s, which is a typical relaxation time for $^1$H. The polarisation is increased to $\approx10\%$ if $T_{1,n}=100$~s, which is relevant to low-$\gamma_n$ nuclei (e.g., $^{15}$N or $^{13}$C) and also relevant for low temperatures. A 10\% polarisation could be obtained for $T_{1,n}=1$~s if the gap is reduced to $h_{\rm cell}\approx1~\mu$m.

An example structure facilitating the implementation of this architecture is depicted in Fig.~\ref{Fig_slab}c. It is composed of two diamond plates structured with $h_{\rm cell}$-wide grooves. With 4~mm~$\times~4$~mm overall lateral dimensions, which corresponds to standard commercially available diamond plates, the sample volume enclosed by this structure would be $\approx 5~\mu$L if the grooves are $\approx200~\mu$m deep. While this volume remains smaller than the capacity of standard NMR probes (100's of $\mu$L), it is already comparable to the capacity of some NMR microprobes and could be increased with deeper grooves, larger diamond plates or by repeating the building block of Fig.~\ref{Fig_slab}c. 

Based on this structure, $h_{\rm cell}=100~\mu$m would be relatively straightforward to realise with standard etching techniques given the aspect ratio close to unity. However, $h_{\rm cell}=1~\mu$m (aspect ratio~$\sim100$) is a much more challenging target that will require further experimental developments. Diamond gratings with aspect ratios of 10-20 are routinely fabricated for optical components~\cite{Forsberg2013,Catalan2016,Li2020}, and diamond needles with aspect ratios up to 50 have also been reported~\cite{Yang2015,Li2018}. We stress that the above requirements correspond to an idealised scenario where the polarisation transfer from NV to sample is optimally efficient. In Sec.~\ref{sec:limitations}, we will discuss practical limitations to this polarisation transfer efficiency, and how these may impose stronger requirements, e.g. a smaller $h_{\rm cell}$.    

We note that hyperpolarisation may add a temporal overhead to the overall NMR acquisition procedure, such that the enhancement of the SNR is reduced compared to the $\langle P \rangle/P_{\rm th}$ ratio~\cite{Lee2014}. This would be the case, for instance, of liquid-state NMR requiring a freeze-thaw process to incorporate the hyperpolarisation step, as will be discussed in Sec.~\ref{sec:diffusion}. However, for solid-state NMR when the hyperpolarisation protocol is applied continuously, the SNR enhancement is essentially given by the polarisation enhancement $\langle P \rangle/P_{\rm th}$. 
 


\subsection{Nanodiamond architecture}

\begin{figure}[b]
	\includegraphics[width=0.49\textwidth]{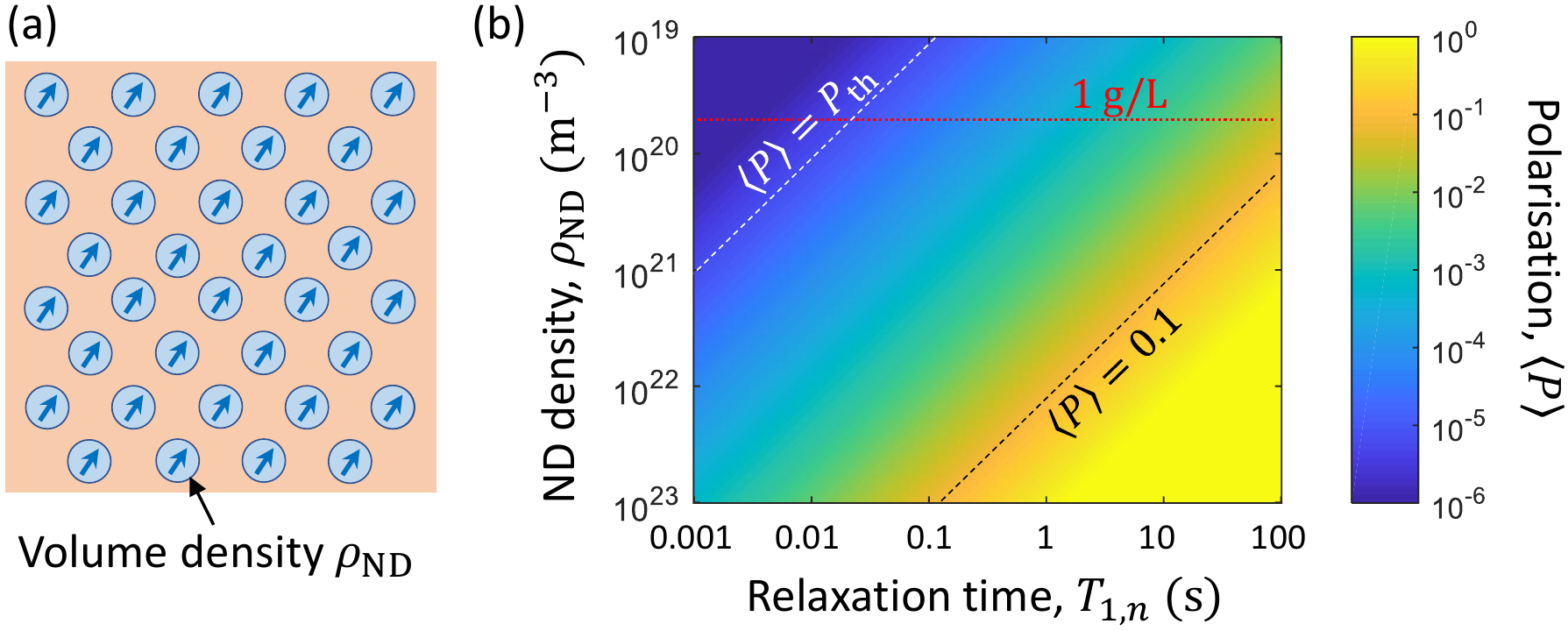}
	\caption{(a) A nanodiamond architecture for the hyperpolarisation of a macroscopic sample (orange region) based on an assembly of spherical nanodiamonds (blue spheres) with a single NV (blue arrows) per nanodiamond. (b) Average polarisation induced by the NVs, $\langle P \rangle$, as a function of $T_{1,n}$ and $\rho_{\rm ND}$, in the geometry (a) for $^1$H spins. The other parameters are: $d_{\rm NV}=10$~nm, $\rho_n=0.6$~nm$^{-3}$. The white (black) dashed line indicates $\langle P \rangle = P_{\rm th}$ ($\langle P \rangle=0.1$). $\langle P \rangle$ is calculated using Eq.~\ref{eq:PNV_ND}, which assumes that the polarisation is $\ll 1$ at any point in the sample. Thus, the plot is not meant to be accurate in the region where $\langle P \rangle$ approaches unity. Where Eq.~\ref{eq:PNV_ND} predicts $\langle P \rangle>1$, the value was capped to 1. The red dotted line indicates an equivalent nanodiamond concentration of 1 g/L. 
	}  
	\label{Fig_ND}
\end{figure}

We next consider the use of spherical nanodiamonds immersed in the sample, with a uniform volume density $\rho_{\rm ND}$ (Fig.~\ref{Fig_ND}a). We take a nanodiamond radius of $d_{\rm NV}=10$~nm, since below this radius NV centres are typically not charge stable on average~\cite{Rondin2010}. We assume that each nanodiamond contains a single NV centre, corresponding to an NV concentration of a few ppm. For simplicity the NV is assumed to be located at the centre of the sphere, with its axis aligned with the external magnetic field. If $N_{\rm s}$ is the number of polarised spins due to a single nanodiamond, the polarisation of the sample will be, on average,
\begin{eqnarray} \label{eq:PNV_ND}
\langle P \rangle & = & \frac{\rho_{\rm ND}N_{\rm s}}{\rho_n} \\
& = & \frac{\rho_{\rm ND}g aT_{1,n}}{\rho_n^{1/2}d_{\rm NV}^{3/2}}
\end{eqnarray} 
where the geometric factor is $g\approx0.29$. 

Figure~\ref{Fig_ND}b plots $\langle P \rangle$ as a function of $T_{1,n}$ and $\rho_{\rm ND}$ for $^1$H spins with a density $\rho_n=0.6$~nm$^{-3}$ (1~M). For $T_{1,n}\approx1$~s, a 10\% polarisation would require  $\rho_{\rm ND}\approx10^{22}$~m$^{-3}$, corresponding to a nanodiamond concentration of $\approx150$~g/L or a $\approx4\%$~volume/volume concentration. Such a concentration is two orders of magnitude larger than that of commercially available colloidal solutions (typically $\sim1$~g/L~\cite{Adamas}) and would be difficult to reach in a liquid environment while maintaining an even dispersion, but could potentially be achieved in a solid mixture.
However, nanodiamonds suffer from significant drawbacks such as a reduced NV charge stability and reduced spin coherence due to the increased surface-to-volume ratio~\cite{Rondin2010,Tetienne2013}, and the random NV orientation implying that only a small proportion of the nanodiamonds would in fact contribute to the polarisation~\cite{Ajoy2018}. For these reasons, the discussions that follow will focus on the slab geometry that appears more promising, although they largely apply to nanodiamonds as well.

\subsection{Practical considerations} \label{sec:limitations}

In this section, we discuss practical aspects of NV experiments relevant to the implementation of hyperpolarisation. First, we will see how the polarisation transfer rate is affected by imperfect NV initialisation, finite NV coherence time, and fast molecular diffusion within the sample. Next, we will discuss the practical limits to increasing the NV density, the effect of parasitic spins, and the requirements associated with laser illumination and background magnetic field.

\subsubsection{NV initialisation} \label{NVinit}

In Sec.~\ref{sec:exchange}, we assumed that the NV electron spin can be perfectly and instantly initialised into a pure spin state. In reality, the NV initialisation fidelity is finite, ${\cal F}_{\rm NV}<1$. A typical observed value is ${\cal F}_{\rm NV}\approx0.8$ for single NV centres in a bulk diamond~\cite{Robledo2011}, but ${\cal F}_{\rm NV}$ may be lower for near-surface NV centres, which are often subject to additional ionisation dynamics~\cite{Bluvstein2019,Dhomkar2018}. Existing strategies to improve this fidelity generally involve significant time overheads and so are not considered here. The factor ${\cal F}_{\rm NV}$ can be simply included in the cooling rate $u({\bf R})$ as a multiplying factor (Eq.~\ref{eq:uR}). 

Furthermore, initialisation of the NV electron spin is not instantaneous. It typically takes a minimum of $1~\mu$s of optical pumping under high laser intensity, followed by $1~\mu$s to allow the NV to relax to its ground state~\cite{Manson2006}, before the polarisation transfer protocol can be applied. This can be taken into account in our model by adding a dead time $t_d\geqslant2~\mu$s to $\tau$ in the denominator of Eq.~\ref{eq:uR}. At lower laser intensity, as often required when addressing large volumes of NV centres (see Sec.~\ref{sec:laser}), this dead time may be as large as 10's of $\mu$s, which could reduce $u({\bf R})$ if the flip-flop time $\tau_0$ is comparable. 

\subsubsection{NV dephasing} \label{NVdephasing}

The NV spin typically experiences some dephasing even in the absence of the sample. Protocols such as PulsePol are, by design, robust against quasistatic dephasing, but faster fluctuations will still contribute to reduce the efficiency of the polarisation transfer. As a crude approximation, this can be taken into account by including a damping factor $e^{-\tau/T_{\rm 2,NV}}$ in the expression of $u({\bf R})$, where $T_{\rm 2,NV}$ is the coherence time of the NV electron spin under the PulsePol sequence (a rigorous treatment of dephasing effects is presented in Ref.~\cite{Hall2020}). For an NV at a depth $d_{\rm NV}=5$~nm, $T_{\rm 2,NV}$ can be as large as $\sim1$~ms with optimised diamond surface preparation and low density of NV centres~\cite{Sangtawesin2019}. However, values in the range $T_{\rm 2,NV}\sim10-100~\mu$s are more commonly observed in samples with large densities of near-surface NVs~\cite{Tetienne2018}. This is much shorter than the optimum interaction time expected for low-$\gamma_n$ nuclei, e.g. $\tau_0\approx1.2$~ms for 1~M of $^{13}$C spins, which will significantly reduce the cooling rate. 

\begin{figure}[b]
	\includegraphics[width=0.35\textwidth]{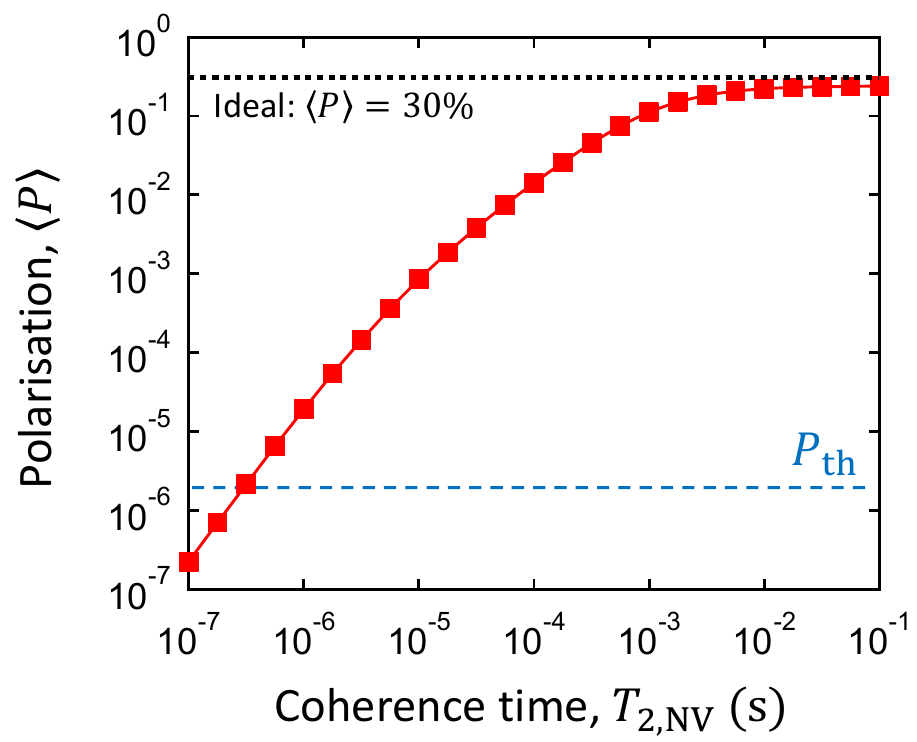}
	\caption{Average polarisation induced by the NVs, $\langle P \rangle$, as a function of the NV spin coherence time, $T_{\rm 2,NV}$. $\langle P \rangle$ is calculated for 1~M of $^{13}$C spins in the geometry of Fig.~\ref{Fig_slab}a with $h_{\rm cell}=10~\mu$m. For each value of $T_{\rm 2,NV}$, the optimum cooling rate from Eq.~\ref{eq:uRextended} is used. The other parameters are: $T_{1,n}=100$~s, $d_{\rm NV}=5$~nm, $\theta_{\rm NV}=54.7^\circ$, ${\cal F}_{\rm NV}=0.8$, $t_d=10~\mu$s, $\sigma_{\rm NV}=10^{16}$~m$^{-2}$. The black solid line corresponds to the ideal case where ${\cal F}_{\rm NV}=1$, $t_d=0$ and $T_{\rm 2,NV}=\infty$. The blue dashed line indicates the thermal polarisation, $P_{\rm th}\approx2.6\times10^{-6}$ assuming $B_0=3$~T and $T=300$~K.        
	}  
	\label{Fig_T2NV}
\end{figure} 

Combining the above factors, we can re-write the cooling rate as
\begin{eqnarray} \label{eq:uRextended}
u({\bf R})=\frac{{\cal F}_{\rm NV}e^{-\frac{\tau}{T_{\rm 2,NV}}}}{\tau+t_d}\frac{A_{\rm s}^2({\bf R})}{A_0^2}\sin^2\left(\frac{A_0\tau}{2}\right).
\end{eqnarray}
The optimum interaction $\tau=\tau_{\rm opt}$ maximising $u({\bf R})$ now depends not only on $A_0$ but also on $t_d$ and $T_{\rm 2,NV}$. The effect of these factors is illustrated in Fig.~\ref{Fig_T2NV}, which plots the average polarisation ($\langle P \rangle$) of 1~M of $^{13}$C spins in the geometry of Fig.~\ref{Fig_slab}a, as a function of $T_{\rm 2,NV}$ assuming ${\cal F}_{\rm NV}=0.8$ and $t_d=10~\mu$s (red line). Compared to the ideal case (horizontal dotted line), $\langle P \rangle$ is reduced from 30\% to 24\% for $T_{\rm 2,NV}\gtrsim10$~ms, and drops to 1.4\% for $T_{\rm 2,NV}=100~\mu$s and $<0.1\%$ for $T_{\rm 2,NV}=10~\mu$s. Even though these polarisation levels still largely exceed the thermal polarisation, it is clear that the finite NV coherence time is an important limiting factor for NV hyperpolarisation. This motivates further work on improving spin coherence of shallow NV centres~\cite{Sangtawesin2019}, which in principle can approach the phonon limit of, e.g., $T_{\rm 2,NV}\sim 1$~s at 77 K~\cite{Bar-Gill2013}, as well as exploring other protocols that may be more efficient in the $T_{\rm 2,NV}$-limited regime~\cite{Hall2020}.

\subsubsection{Molecular diffusion} \label{sec:diffusion}
         
Most conventional DNP methods are inefficient in the presence of fast molecular diffusion, except for the Overhauser effect in some conditions~\cite{Ravera2016}. For NV centres located several nanometres from the surface, however, the interaction correlation time is too long for the Overhauser effect to work effectively~\cite{Chen2016}, but is still too short for other methods including the coherent polarisation exchange scenario considered in Sec.~\ref{sec:exchange}. 

To see that, consider the example of liquid water at room temperature, which has a diffusion coefficient of $D_n\approx2.5\times10^{-9}$~m$^2$s$^{-1}$. Over the time scale of the flip-flop dynamics ($\tau_0\approx30~\mu$s), the $^1$H spins travel over typical distances $\sim\sqrt{D_n\tau_0}\sim300$~nm. This is much larger than the distance over which the NV-nuclear coupling is significant (of the order of $d_{\rm NV}$). In the framework of Sec.~\ref{sec:exchange}, it is as if the state of the nuclear spins was being reset to an unpolarised, incoherent mixture every $t_D\sim d_{\rm NV}^2/D_n=10$~ns for $d_{\rm NV}=5$~nm. This leads to a Zeno-type effect whereby the NV spin, instead of exchanging polarisation following a $\sin^2\left(\frac{A_0\tau}{2}\right)$ law, would supply an amount $\left[ \sin^2\left(\frac{A_0 t_D}{2}\right) \right]^{\frac{\tau}{t_D}}\approx \left(\frac{A_0 t_D}{2}\right)^{\frac{2\tau}{t_D}}$, which becomes rapidly negligible for $t_D\ll\tau_0$. 

Thus, the results presented in the previous sections are valid only for solid samples or high-viscosity liquid samples, typically $D_n\lesssim10^{-14}$~m$^2$s$^{-1}$. 
Nevertheless, liquid samples such as aqueous solutions could be handled in a similar fashion to conventional dissolution-DNP, where the solution is frozen for the hyperpolarisation step and thawed for the NMR measurement~\cite{Larsen2003,Joo2006,Kouril2019}. Here, the solution would simply need to be cooled below the freezing point, in contrast to DNP where a much lower temperature is required to achieve high polarisation.


\subsubsection{NV density} \label{sec:density}

In Sec.~\ref{sec:slab}, we introduced a surface density of NV centres, $\sigma_{\rm NV}$, which should be maximised to increase the average polarisation, according to Eq.~\ref{eq:PNV_slab}. Here, we discuss the practical limits to increasing $\sigma_{\rm NV}$.

Dense ensembles of near-surface NV centres are typically produced by nitrogen (N) ion implantation~\cite{Tetienne2018}. For a fluence of $10^{13}$~N/cm$^2$, assuming a 4\% N-to-NV conversion efficiency and taking into account that only 25\% of the NV centres will have the correct crystallographic orientation and be aligned with the applied magnetic field~\cite{Pham2012}, the surface density of ``active'' NVs is $\sigma_{\rm NV}=10^{15}$~m$^{-2}$. Obtaining larger surface densities with this method is challenging especially at the low implantation energies required to create shallow NV centres, e.g. 2.5 keV to obtain a depth $d_{\rm NV}\approx5$~nm. Indeed, a 2.5 keV implant with a $10^{13}$~N/cm$^2$ fluence already creates locally about 100 ppm of N, and 2000 ppm of vacancies before annealing, according to Stopping and Range of Ions in Matter (SRIM) Monte Carlo simulations. Significantly larger fluences would likely cause irreparable damage and significantly reduce the spin coherence time $T_{\rm 2,NV}$ (limited by the bath of surrounding paramagnetic impurities). It might be possible to increase the NV density without increasing the fluence, through doping engineering to improve the N-to-NV yield~\cite{DeOliveira2017,Herbschleb2019,Luhmann2019}, but further work is needed to test the efficiency of this approach for dense layers of shallow NVs.  

However, for applications where the NVs are not used for readout as in Sec.~\ref{sec:slab}, the NVs do not need to be restricted to the near-surface region, which opens the possibility to use optimised bulk doping techniques. In particular, a record high bulk NV density of 45 ppm was achieved in Ref.~\cite{Kucsko2018}, through electron irradiation and in-situ annealing of a diamond naturally containing about 100 ppm of nitrogen. Counting only those 25\% of the NVs that would be aligned with the external magnetic field, i.e. $\approx11$~ppm of active NVs, a 5-nm slice in this diamond would give a surface density $\sigma_{\rm NV}\approx 10^{16}$~m$^{-2}$. In Sec.~\ref{sec:slab}, we assumed $\sigma_{\rm NV}=10^{16}$~m$^{-2}$ with a fixed depth $d_{\rm NV}=5$~nm, which is therefore a good approximation for this diamond. 

It is important to note that the NV density directly competes with the NV coherence time $T_{\rm 2,NV}$. For instance, when substitutional nitrogen (N) is the dominant impurity, the NV coherence time is inversely proportional to the density [N]~\cite{Bauch2019}. Since the NV density is proportional to [N], assuming a constant conversion efficiency, increasing the NV density therefore does not necessarily result in an increased polarisation. Dynamical decoupling experiments performed on diamonds with [N]~$\sim100$~ppm reported coherence times of up to $T_{\rm 2,NV}\sim 100~\mu$s~\cite{DeLange2010}. This is sufficient not to limit the cooling rate in the case of a dense $^1$H ensemble, but is already a limiting factor for more dilute samples (see Fig.~\ref{Fig_T2NV}). Thus, the assumed density of $\sigma_{\rm NV}=10^{16}$~m$^{-2}$ can be considered an optimum trade-off with respect to $T_{\rm 2,NV}$.

We note that a surface density $\sigma_{\rm NV}=10^{16}$~m$^{-2}$ corresponds to a typical lateral distance between NVs of $\approx10$~nm. Given that the polarisation exchange dynamics between NV and nuclear spins is dominated by the most strongly coupled nuclear spins at a distance $\sim d-2d$, the presence of nearby NV centres in the array may affect this dynamics slightly, and as a result change the number of polarised spins per NV, $N_{\rm s}$. For simplicity, this potential correction was neglected in Sec.~\ref{sec:slab}.

\subsubsection{Parasitic spins} \label{sec:parasitic}

In addition to the NV spins, the diamond hosts a number of parasitic spins that may contribute to reducing the amount of polarisation reaching the target sample, via two different mechanisms. First, unpaired electron spins from defects located inside the diamond (such as nitrogen impurities or vacancy clusters~\cite{Tetienne2018}) or on the diamond surface~\cite{Grinolds2014,Rosskopf2014,Myers2014,Romach2015} act as a source of dephasing for the nuclear spins in the sample, which could affect the dynamics of the NV-sample system during the polarisation transfer and therefore the efficiency of the process~\cite{Hall2020}. 

Second, the diamond hosts nuclear spins (inside or at the surface) that could act as competing polarisation sinks reducing the amount of polarisation transferred to the target sample. Inside the diamond, $^{14}$N and/or $^{15}$N spins are typically present, as well as $^{13}$C spins if the diamond is not isotopically purified. Moreover, an adventitious layer of $^1$H spins of thickness 1-2~nm (density $\rho_n\sim50$~nm$^{-3}$) is often observed on the diamond surface~\cite{DeVience2014,Staudacher2015}. 

\subsubsection{Laser illumination} \label{sec:laser}

An important element of NV experiments is the laser illumination at 532 nm wavelength (or similar), which enables the initialisation of the NV spin to a nearly pure state, as discussed in Sec.~\ref{NVinit}. The illumination time required to fully initialise the NV spin depends on the laser intensity, with $\sim100$~kW/cm$^2$ typically employed to initialise the NV in $\sim1~\mu$s~\cite{Manson2006}. To initialise a 1~mm$^2$ array of NVs at normal incidence, a peak laser power of 1 kW would be required to achieve this initialisation time, highlighting the enormous challenge posed by laser illumination in mm-sized devices such as the structure presented in Fig.~\ref{Fig_slab}c. While lower laser intensities may be used, this comes at the cost of a reduced polarisation rate (see Sec.~\ref{NVinit}). Careful optical engineering exploiting, e.g., waveguiding and multi-pass strategies~\cite{Clevenson2015}, will therefore be critical for this application.

A deleterious consequence of laser illumination is the heating or photo-damage it may induce on the sample to be polarised. Laser-induced heating is particularly problematic when the sample must be frozen to allow polarisation. Critically, the desired close proximity between NV and sample ($d_{\rm NV}\sim5$~nm) means that the laser intensity in the sample (near the diamond surface) will be similar to that in the NV layer regardless of the illumination configuration, and precludes the use of laser shielding. However, we note that laser heating, if it can be tuned appropriately, could also be used as a resource to thaw the sample after it has been frozen for the polarisation step, a solution exploited in previous DNP experiments~\cite{Joo2006}.  

\subsubsection{External magnetic field}
     
In principle, the class of protocols considered here, such as NOVEL or PulsePol, can operate at low magnetic field ($B_0\lesssim0.1$~T) as well as higher magnetic field (e.g. $B_0\sim5$~T), although the high field regime is substantially more technically demanding, requiring a high-power, high-frequency microwave source; similar to most conventional DNP methods.

The ability to operate at low fields therefore makes NV hyperpolarisation potentially simpler and more cost-effective than DNP. However, the subsequent measurement of the hyperpolarised sample in a high-field NMR spectrometer comes with its own challenges, and care must be taken to minimise polarisation losses~\cite{Ajoy2019}. 

One requirement specific to the NV system is the need for the magnetic field to be aligned with the symmetry axis of the NV centres (within a few degrees), to ensure effective spin polarisation by optical pumping~\cite{Tetienne2012}. The diamond hyperpolariser must therefore be designed to facilitate this alignment, for instance (111)-oriented diamond surfaces (i.e. $\theta_{\rm NV}=0$ in Fig.~\ref{Fig_definitions}b) may prove easier to accommodate. Conveniently, $\theta_{\rm NV}=0$ is the angle that gives the largest cooling rate (Fig.~\ref{Fig_Neff_ideal}b).

\section{NV-based hyperpolarisation for NV-detected NMR} \label{sec:NV-NMR}
  
We now turn our attention to the situation where the NMR signal is detected using a local magnetometer such as an NV centre in the same diamond as that used for hyperpolarisation, rather than with a remote inductive detector as in a conventional NMR spectrometer. The motivation is that the local polarisation may be significantly larger than the average polarisation, which suggests that the NMR signal detected by the NVs could be enhanced compared to the signal obtained in the absence of hyperpolarisation. 

Two scenarios are analysed. In the first one, a layer of near-surface NV centres is used to generate the hyperpolarisation but the NMR signal is detected by NV centres located deeper in the diamond, typically several $\mu$m from the surface. This scenario, referred to as  micro-NMR, could enable liquid-state NMR spectroscopy with a sensitivity and spectral resolution approaching that of conventional NMR, but with a greatly simplified apparatus. In the second scenario, referred to as nano-NMR, the same near-surface NV centres are used both for the hyperpolarisation and for the NMR detection, which could find applications in NMR studies of nanoscale objects or NMR imaging with a sub-$\mu$m spatial resolution.

\subsection{Micro-NMR} \label{sec:micro-NMR}

\subsubsection{Background}

In NV-based micro-NMR, as first demonstrated in Ref.~\cite{Glenn2018}, the NMR signal is read out by NV centres located several $\mu$m away from the sample ($d_{\rm RO}$). The probe distance $d_{\rm RO}$ is chosen large enough so that diffusion of the nuclear spins in and out of the sensing volume (of size roughly given by $d_{\rm RO}$) does not limit the NMR spectral resolution. For a liquid sample with diffusion coefficient $D_n$, the interaction correlation time due to translational diffusion is $\tau_c\approx\frac{2d_{\rm RO}^2}{D_n}$, which must be larger than the dephasing time of the nuclear spins, $T_{2,n}^*$, in order to avoid spectral broadening~\cite{Pham2016}. For typical aqueous solutions, this implies $d_{\rm RO}\gtrsim10~\mu$m~\cite{Glenn2018,Smits2019}.

\begin{figure}[b]
	\includegraphics[width=0.5\textwidth]{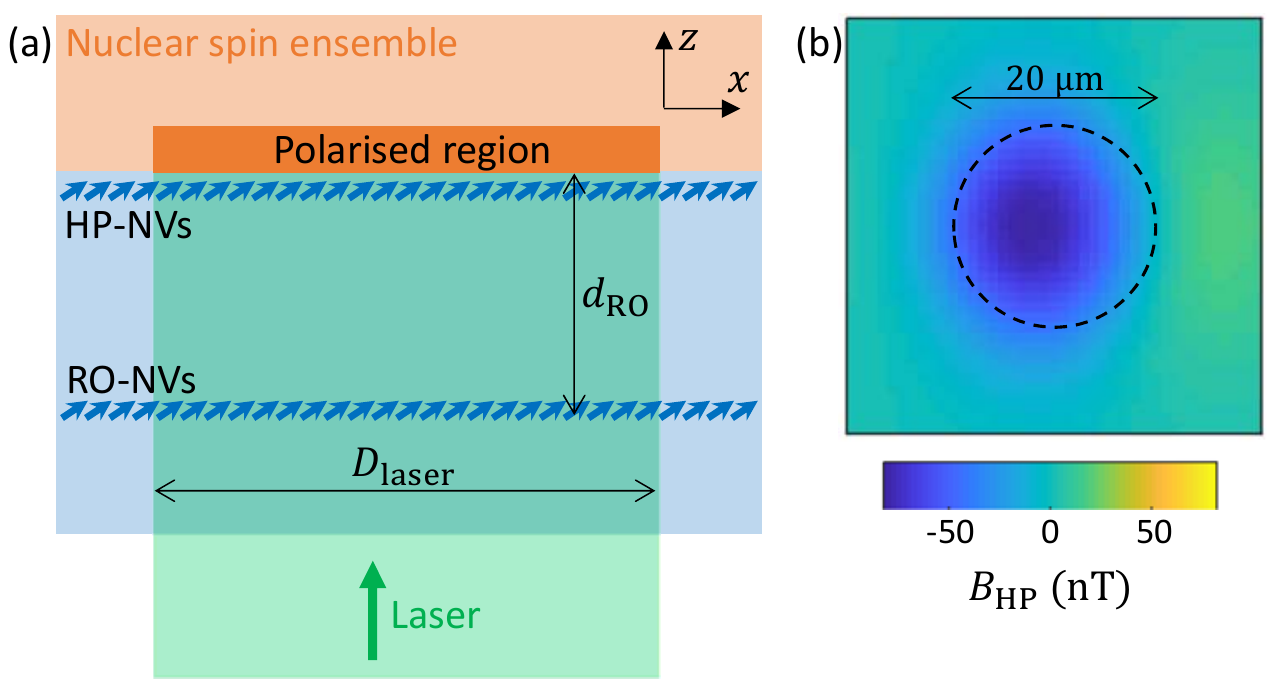}
	\caption{(a) Proposed set-up to combine liquid-state micro-NMR (using a layer of deep readout NVs, or RO-NVs) with in-situ NV hyperpolarisation (shallow NVs, or HP-NVs). (b) Calculated magnetic field amplitude $B_{\rm HP}$ produced by the polarised region in the plane of the RO-NVs for $d_{\rm RO}=10~\mu$m. The dashed circle represents the laser beam of diameter $D_{\rm laser}=20~\mu$m. The magnetisation $\tilde{\bf M}_{\rm HP}$ was calculated from Eq.~\ref{eq:MHP} assuming a density of HP-NVs $\sigma_{\rm HP}=10^{16}$~m$^{-2}$ and a number of polarised spins per HP-NV of $N_{\rm s}\approx37,000$ corresponding to the steady state polarisation obtained from Sec.~\ref{sec:singleNV} with the following parameters: $\rho_n=66$~nm$^{-3}$, $d_{\rm NV}=5$~nm, $\theta_{\rm NV}=54.7^\circ$, $T_{1,n}=1$~s.	 
	}  
	\label{Fig_NV-NMR_model}
\end{figure}   

NV-based NMR is generally conducted in a low magnetic field for experimental convenience. Since the NMR signal is proportional to the thermal polarisation in this case, there is ample room for boosting the signal and hence the sensitivity by applying hyperpolarisation techniques. For instance, the polarisation was only $P_{\rm th}\approx3\times10^{-7}$ in the original demonstration~\cite{Glenn2018} ($B_0=88$~mT), which was later increased by more than two orders of magnitude by in-situ liquid-state DNP based on the Overhauser effect~\cite{Bucher2020}. Here, we analyse the possibility of using NV centres to hyperpolarise the sample in situ, which would have the advantage of not having to introduce free radicals to the analyte. 

A major drawback, however, is that NV hyperpolarisation is not applicable to liquid samples, as we discussed in Sec.~\ref{sec:diffusion}. One could envision increasing the viscosity of the solution, but $T_{2,n}^*$ would then decrease due to dipolar broadening, deteriorating the NMR linewidth. Moreover, in this case the distance $d_{\rm RO}$ could be reduced without compromising the spectral resolution, down to a few nanometres for $D_n\lesssim10^{-14}$~m$^2$s$^{-1}$. This corresponds to the nano-NMR regime analysed in Sec.~\ref{sec:nanoNMR}. Thus, in order to combine NV hyperpolarisation to liquid-state micro-NMR, it is necessary to freeze the solution for the polarisation step (in principle, at any temperature below the freezing point) and then thaw it before the NMR measurement, as in dissolution-DNP~\cite{Larsen2003,Joo2006,Kouril2019}. 

\subsubsection{Model}

Let us consider the configuration of Fig.~\ref{Fig_NV-NMR_model}a. An array of near-surface NV centres (referred to as the HP-NVs), with a depth $d_{\rm NV}\sim5$~nm and surface density $\sigma_{\rm HP}$, is used to polarise the frozen-solution sample by applying the protocol described in Sec.~\ref{sec:singleNV}. Once the polarisation has reached saturation, i.e. after a time $\sim T_{1,n}$, the sample is rapidly thawed and the NMR measurement can proceed using an array of readout NV centres (referred to as the RO-NVs) at a depth $d_{\rm RO}$. We assume that the two NV arrays are excited by the same laser beam of diameter $D_{\rm laser}$, with a typical value $D_{\rm laser}\approx20~\mu$m~\cite{Glenn2018}. 

The polarisation step generates a disk-like region of polarised sample of diameter $\approx D_{\rm laser}$. Because the probe distance $d_{\rm RO}$ is much larger than the extent of the  polarisation region from a single HP-NV ($\sim d_{\rm NV}$) and than the lateral separation between HP-NVs ($\sim\sigma_{\rm HP}^{-1/2}$), the stray field seen by the RO-NVs can be calculated by approximating the total polarised region as a thin disk of uniform magnetisation. That is, we average out the spatial variations of the polarisation over length scales much smaller than $d_{\rm RO}$. The areal magnetisation of this disk is then simply 
\begin{eqnarray} \label{eq:MHP}
\tilde{\bf M}_{\rm HP}=\sigma_{\rm HP}N_{\rm s}{\bf m}_n
\end{eqnarray}
where $N_{\rm s}$ is the effective number of polarised spins due to a single HP-NV (as calculated in Sec.~\ref{sec:buildup}) and ${\bf m}_n$ is the magnetic moment of a single nuclear spin. The tilde in $\tilde{\bf M}_{\rm HP}$ denotes the fact that this is an areal magnetisation (in units of A) rather than a volume magnetisation (in A/m). 

Upon thawing, the polarisation will diffuse both laterally and vertically (away from the diamond surface). However, the spatial extent of this diffusion should be confined to a volume of the order of $d_{\rm RO}$ by the time the NMR measurement is completed, since this was the selection criterion for $d_{\rm RO}$. One could also imagine to use the laser beam to induce the thawing process, in which case it might even be possible to keep a frozen containment structure surrounding the liquid core. In any case, we will assume for simplicity that the shape of the polarised region is approximately preserved upon thawing and throughout the NMR measurement. 

Following an RF $\pi/2$ pulse, the magnetisation precesses about the NV axis (unit vector ${\bf u}_{\rm NV}$), generating an AC magnetic field of amplitude $B_{\rm HP}$ at the position of a given RO-NV. $B_{\rm HP}$ is the projection of the field along the quantization axis of the RO-NV, i.e. ${\bf u}_{\rm NV}$. To calculate $B_{\rm HP}$, we compute the stray field generated by a magnetic disk of magnetisation $\tilde{\bf M}_{\rm HP}$ given by Eq.~\ref{eq:MHP}, where we choose the direction of ${\bf m}_n$ that maximises the field seen by the RO-NVs. For instance, if ${\bf u}_{\rm NV}$ lies in the $xz$ plane, i.e. ${\bf u}_{\rm NV}=(\sin\theta_{\rm NV},0,\cos\theta_{\rm NV})$ in Cartesian coordinates, we take ${\bf m}_n=\frac{\hbar\gamma_n}{2}{\bf u}_{\rm y}\times {\bf u}_{\rm NV}=\frac{\hbar\gamma_n}{2}(\cos\theta_{\rm NV},0,-\sin\theta_{\rm NV})$. 
     
\subsubsection{Results}     
     
As an example, Fig.~\ref{Fig_NV-NMR_model}b shows the calculated $B_{\rm HP}$ amplitude in the plane of the RO-NVs at $d_{\rm RO}=10~\mu$m, from a polarised region of diameter $D_{\rm laser}=20~\mu$m, with an NV angle $\theta_{\rm NV}=54.7^\circ$. In these simulations, the HP-NV layer is characterised by $d_{\rm NV}=5$~nm and $\sigma_{\rm HP}=10^{16}$~m$^{-2}$, and the sample mimics frozen water ($^1$H spins with $\rho_n=66$~nm$^{-3}$ and $T_{1,n}=1$~s). Averaging $B_{\rm HP}$ over the readout disk (dashed circle in Fig.~\ref{Fig_NV-NMR_model}b), we obtain $\langle B_{\rm HP}\rangle\approx42$~nT in this case.  


\begin{figure}[t]
	\includegraphics[width=0.48\textwidth]{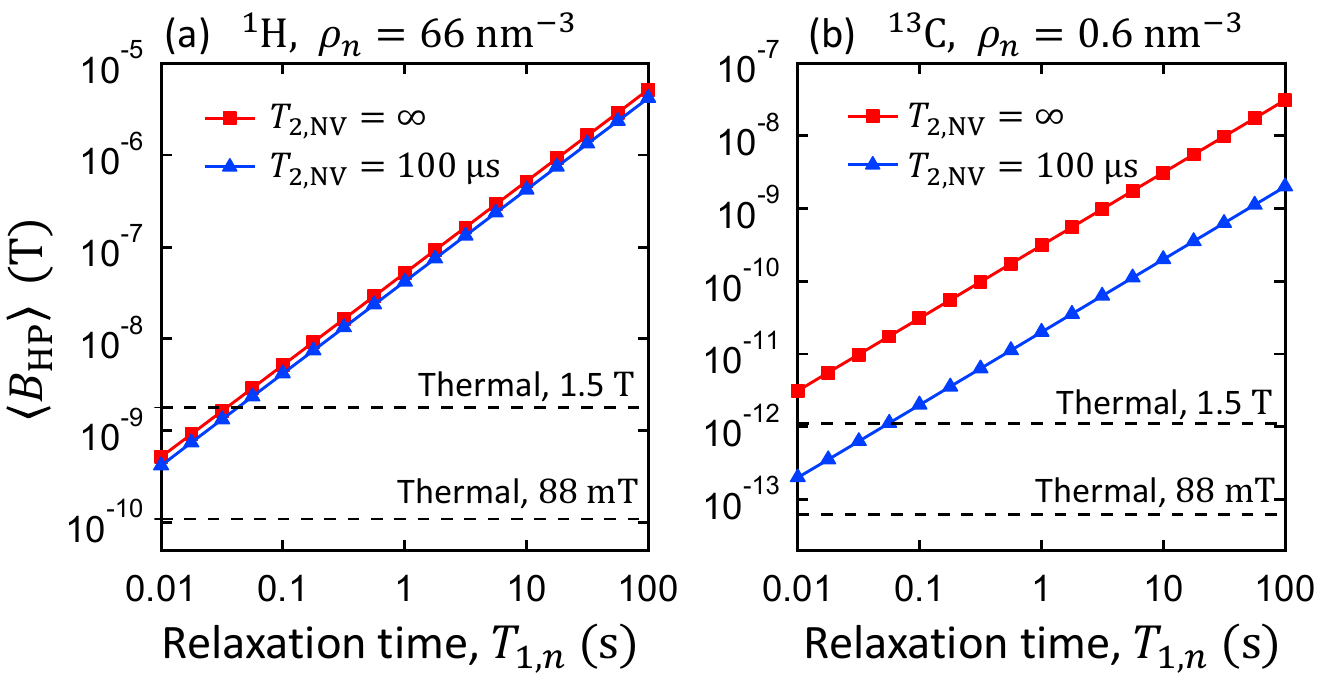}
	\caption{(a,b) Average magnetic field seen by the RO-NVs as a result of hyperpolarisation by the HP-NVs, $\langle B_{\rm HP}\rangle$, plotted as a function of the $T_{1,n}$ time of the target spins in the frozen phase. The sample is $\rho_n=66$~nm$^{-3}$ of $^1$H spins in (a) and 1~M (or $0.6$~nm$^{-3}$) of $^{13}$C spins in (b). The magnetisation $\tilde{\bf M}_{\rm HP}$ is calculated assuming $\sigma_{\rm HP}=10^{16}$~m$^{-2}$, $d_{\rm NV}=5$~nm, $\theta_{\rm NV}=54.7^\circ$ and $T_{2,\rm NV}=\infty$ (red data) or $T_{2,\rm NV}=100~\mu$s (blue). The stray field $\langle B_{\rm HP}\rangle$ is calculated assuming $d_{\rm RO}=10~\mu$m and $D_{\rm laser}=20~\mu$m. The dashed lines indicate the field amplitude seen by the RO-NVs due to thermal polarisation only, at a field $B_0=1.5$~T or $B_0=88$~mT.  
	}  
	\label{Fig_NV-NMR_ratio}
\end{figure}  

The average field $\langle B_{\rm HP}\rangle$ is proportional to the amplitude of $\tilde{\bf M}_{\rm HP}$ hence scales as $\langle B_{\rm HP}\rangle\propto \sigma_{\rm HP}\gamma_n\rho_n^{1/2}d_{\rm NV}^{-3/2}T_{1,n}$, where $T_{1,n}$ is the longitudinal spin relaxation time in the frozen phase. To illustrate the range of fields one could obtain, Fig.~\ref{Fig_NV-NMR_ratio} plots $\langle B_{\rm HP}\rangle$ as a function of $T_{1,n}$ for a solution with a high density of $^1$H spins (Fig.~\ref{Fig_NV-NMR_ratio}a) and for 1~M of $^{13}$C spins (Fig.~\ref{Fig_NV-NMR_ratio}b). The hyperpolarisation step assumes $\sigma_{\rm HP}=10^{16}$~m$^{-2}$, $d_{\rm NV}=5$~nm and $T_{2,\rm NV}=\infty$ (red data) or $T_{2,\rm NV}=100~\mu$s (blue).

The values obtained can be compared to the field amplitude $B_{\rm th}$ that would be obtained from thermal polarisation only. The volume magnetisation is
\begin{eqnarray} \label{eq:Mth}
{\bf M}_{\rm th}=\rho_n P_{\rm th} {\bf m}_n
\end{eqnarray}
where $P_{\rm th}$ is given by Eq.~\ref{eq:Pth}. In general, the stray field $B_{\rm th}$ depends on the portion of sample contributing to the magnetisation ${\bf M}_{\rm th}$, which depends on technical details such as the homogeneity of the RF driving field. Nevertheless, if the corresponding volume has a size much larger than $d_{\rm RO}$ and $D_{\rm laser}$, as is typically the case, then $B_{\rm th}$ is uniform in the RO-NV plane and depends little on the exact volume and shape of the active part of the sample~\cite{Smits2019}. For simplicity, we will therefore consider a sample of cubic shape, for which the stray field can be calculated analytically. The resulting field is indicated by dashed lines in Fig.~\ref{Fig_NV-NMR_ratio}a,b, where we show the cases $B_0=88$~mT (similar to Ref.~\cite{Glenn2018}) and $B_0=1.5$~T (similar to the pre-polarisation stage in Ref.~\cite{Smits2019}).

For the high density $^1$H solution (Fig.~\ref{Fig_NV-NMR_ratio}a), we see that with a modest value of $T_{1,n}=1$~s, $\langle B_{\rm HP}\rangle$ exceeds $B_{\rm th}$ by a factor $\approx400$ at 88~mT, and by a factor $\approx20$ at 1.5~T, with a negligible reduction caused by a finite $T_{2,\rm NV}$.  Assuming the freeze-thaw process can be repeated many times, with a hyperpolarisation time ($\sim T_{1,n}$) of the same order as the measurement time (assuming $T_{2,n}^*\sim T_{1,n}$ which is plausible for a liquid), the temporal overhead could be just a few seconds (including freezing and thawing times) for $\sim1$~s of measurement. In this case, the signal enhancement $\langle B_{\rm HP}\rangle/B_{\rm th}$ would translate into a comparable level of SNR enhancement, suggesting that NV hyperpolarisation could be viable in this scenario. 

A relatively straightforward improvement could come from structuring the diamond surface to form a nanograting, as demonstrated in Ref.~\cite{Kehayias2017}. This would effectively increase the NV surface density $\sigma_{\rm HP}$, translating into an increase in $\langle B_{\rm HP}\rangle$ by the same amount. A 15-fold increase in $\sigma_{\rm HP}$ was demonstrated in Ref.~\cite{Kehayias2017}, and even larger enhancements can be anticipated with higher aspect ratios~\cite{Forsberg2013,Catalan2016,Li2020,Yang2015,Li2018}.     

For a solution containing 1~M of $^{13}$C spins (Fig.~\ref{Fig_NV-NMR_ratio}b), the enhancement $\langle B_{\rm HP}\rangle/B_{\rm th}$ is larger in the ideal case, but is reduced by an order of magnitude when taking into account a $T_{2,\rm NV}=100~\mu$s, which is due to the fact that the flip-flop time is now longer ($\tau_0\approx1.2$~ms). Nevertheless, with this finite $T_{2,\rm NV}=100~\mu$s we still predict a 300-fold enhancement over $B_{\rm th}$ at 88~mT with $T_{1,n}=1$~s.

\subsection{Nano-NMR} \label{sec:nanoNMR} 

\subsubsection{Background} 

We now consider the possibility of using NV hyperpolarisation to benefit nano-NMR or nano-MRI experiments.
Here, we refer to the use of a single near-surface NV centre (typically, $d_{\rm NV}\lesssim10$~nm) to perform NMR spectroscopy on a nanoscale volume of order $d_{\rm NV}^3$~\cite{Staudacher2013,Mamin2013,Lovchinsky2016,Aslam2017} (``nano-NMR''), or the use of a dense two-dimensional layer of near-surface NV centres to image the nano-NMR signal on a camera~\cite{DeVience2014,Ziem2019} (``nano-MRI''). Nano-NMR may be useful to study surface interactions and the dynamics of molecules at the nanoscale through the analysis of correlations in the NMR signal~\cite{Staudacher2015} or to characterise interactions in nanoscale materials~\cite{Lovchinsky2017}, whereas nano-MRI allows characterisation of samples over larger scales, with a sub-micrometer lateral spatial resolution (limited by optical diffraction). 

Previous demonstrations of nano-NMR/MRI have used the NVs to detect the statistical polarisation of the nuclear spins~\cite{Staudacher2013,Mamin2013,Lovchinsky2016,Aslam2017}, as it is much larger than the thermal (Boltzmann) polarisation for nanoscale volumes. Statistical polarisation spontaneously generates a magnetic field oscillating at the Larmor frequency of the nuclear spins. Because the phase of this oscillating field is random, time-averaged NV measurements are sensitive to the variance of this field, $B_{\rm rms}^2$, which is given by~\cite{Pham2016}
\begin{eqnarray} \label{eq:Brms}
B_{\rm rms}^2 = \rho_n\left( \frac{\mu_0\hbar\gamma_n}{4\pi} \right)^2 \frac{\pi[8-3\sin^4(\theta_{\rm NV})]}{128d_{\rm NV}^3}
\end{eqnarray} 
in the geometry of Fig.~\ref{Fig_definitions}b.

While there are several ways to generate an NMR spectrum~\cite{Staudacher2013,Mamin2013,Laraoui2013}, they all produce a measurable ``signal'' (namely, a change in NV spin population) of the order of $\Delta p_{\rm NV}^{\rm rms}\sim(\gamma_e B_{\rm rms}\tau_{\rm sens})^2$ in the small signal limit, where $\tau_{\rm sens}$ is the interrogation (sensing) time, which is limited by the NV coherence time $T_{\rm 2,NV}$. For a dense ensemble of $^1$H spins ($\rho_n=66$~nm$^{-3}$), with an NV at $d_{\rm NV}=5$~nm and $\theta_{\rm NV}=54.7^\circ$, we obtain $B_{\rm rms}\approx830$~nT, giving a nearly full contrast ($\Delta p_{\rm NV}^{\rm rms}\sim 1$) in only $\tau_{\rm sens}\sim 10~\mu$s, which is typically well within $T_{\rm 2,NV}$. For a more dilute sample ($\rho_n=0.6$~nm$^{-3}$), however, $B_{\rm rms}\approx80$~nT which gives only $\Delta p_{\rm NV}^{\rm rms}\sim10^{-2}$ with $\tau_{\rm sens}\sim10~\mu$s. 


On the other hand, NV hyperpolarisation followed by a $\pi/2$ RF pulse on the nuclear spins would generate a signal of the form $\Delta p_{\rm NV}^{\rm HP}\sim\gamma_e B_{\rm HP}\tau_{\rm sens}$, where $B_{\rm HP}$ is the amplitude of the AC magnetic field from the polarised spins, evaluated at the NV location~\cite{deLange2011}. This amplitude can be computed as the sum of the dipolar field from each nuclear spin, projected along the NV axis,
\begin{eqnarray} \label{eq:Bac}
B_{\rm HP} = \rho_n \int P({\bf R},\infty){\bf b}_{\rm AC}({\bf R})  \cdot {\bf u}_{\rm NV}~{\rm d}^3{\bf R}
\end{eqnarray}
with
\begin{eqnarray} \label{eq:bac}
{\bf b}_{\rm AC}({\bf R}) = \frac{\mu_0}{4\pi} \left( \frac{3({\bf m}_n\cdot{\bf R}){\bf R}}{R^5}-\frac{{\bf m}_n}{R^3} \right),
\end{eqnarray}
where ${\bf u}_{\rm NV}$ is the NV axis unit vector, ${\bf m}_n$ is the magnetic moment of the nuclear spins in the transverse position that maximises $B_{\rm HP}$, and $P({\bf R},\infty)$ is the steady-state polarisation distribution, solution of Eq.~\ref{eq:master}. 

In nano-MRI applications where a dense layer of NV centres is addressed at once, the polarisation distribution should include the effect of all the NVs in the array. However, we found that the dominant contribution to $B_{\rm HP}$ at a given NV site comes from the polarisation imparted by this same NV, with a negligible contribution from the neighbouring NVs even at NV densities as large as $\sigma_{\rm HP}=10^{16}$~m$^{-2}$. Therefore, for generality the results below present the single NV limit.

\subsubsection{Results} 

\begin{figure}[t]
	\includegraphics[width=0.5\textwidth]{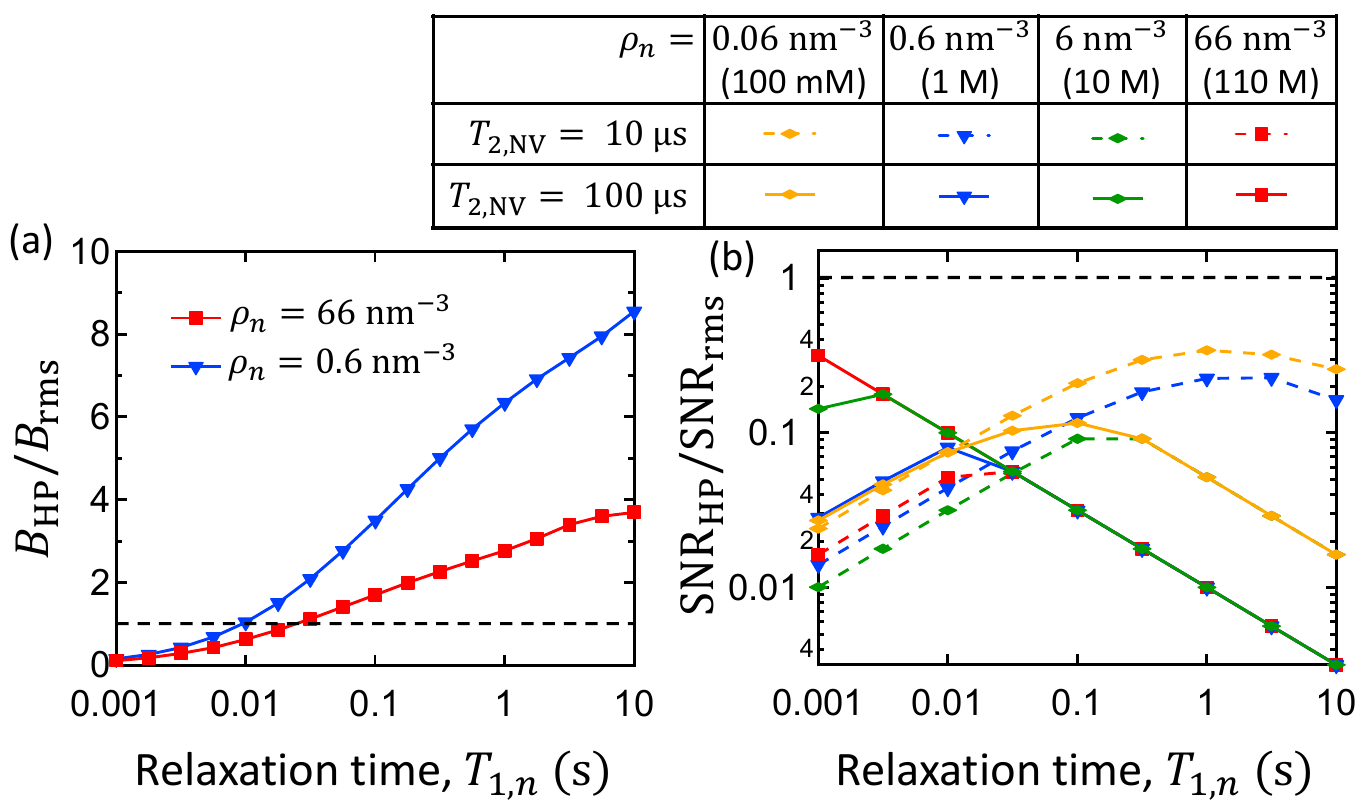}
	\caption{(a) Calculated ratio $B_{\rm HP}/B_{\rm rms}$ as a function of $T_{1,n}$ for $^1$H spins with $\rho_n=66$~nm$^{-3}$ (red data, $B_{\rm rms}\approx830$~nT) and $\rho_n=0.6$~nm$^{-3}$ (blue, $B_{\rm rms}\approx80$~nT). Diffusion is included with a coefficient $D_n=0.22\frac{\mu_0}{4\pi}\hbar\gamma_n^2\rho_n^{1/3}$. Other parameters are: $d_{\rm NV}=5$~nm, $\theta_{\rm NV}=54.7^\circ$, $T_{\rm 2,NV}=1$~ms. (b) Calculated ratio ${\rm SNR}_{\rm HP}/{\rm SNR}_{\rm rms}$ (from Eq.~\ref{eq:SNR2}) as a function of $T_{1,n}$ for $^1$H spins with $\rho_n=66$~nm$^{-3}$ (red data), $\rho_n=6$~nm$^{-3}$ (green), $\rho_n=0.6$~nm$^{-3}$ (blue) and $\rho_n=0.06$~nm$^{-3}$ (orange). The NV coherence time is $T_{\rm 2,NV}=10~\mu$s (dashed lines) or $100~\mu$s (solid lines). The correlation time of the AC field is taken to be $\tau_c=100~\mu$s. Other parameters are as in (a). 
	}  
	\label{Fig_nanoNMR}
\end{figure}

To compare the amplitude $B_{\rm HP}$ obtained from NV hyperpolarisation with $B_{\rm rms}$, we plot the ratio $B_{\rm HP}/B_{\rm rms}$ as a function of $T_{1,n}$ for a semi-infinite ensemble of $^1$H spins (Fig.~\ref{Fig_nanoNMR}a). Two densities are compared, $\rho_n=66$~nm$^{-3}$ (red data) and $\rho_n=0.6$~nm$^{-3}$ (blue), and a diffusion coefficient corresponding to the solid-state case, $D_n\approx0.22\frac{\mu_0}{4\pi}\hbar\gamma_n^2\rho_n^{1/3}$~\cite{Cheung1981}, is included. The other parameters are taken as $d_{\rm NV}=5$~nm, $\theta_{\rm NV}=54.7^\circ$, and $T_{\rm 2,NV}=1$~ms. The ratio $B_{\rm HP}/B_{\rm rms}$ is found to vary relatively weakly with $T_{1,n}$, in a roughly logarithmic manner. We have $B_{\rm HP}/B_{\rm rms}\sim1$ for $T_{1,n}=10$~ms, increasing to $B_{\rm HP}/B_{\rm rms}\approx8$ for the dilute sample when $T_{1,n}=10$~s and only $B_{\rm HP}/B_{\rm rms}\approx4$ for the dense sample. This weak dependence can be understood by considering the role of diffusion: even though the total number of polarised spins $N_{\rm s}$ increases linearly with $T_{1,n}$ (Eq.~\ref{eq:Neff_scaling}), this polarisation extends further away from the NV spin on average thus causing a comparatively small increase in the local magnetic field seen by the NV; this effect is more marked in the dense sample when diffusion is faster.

We now examine how the SNR would change in an experiment detecting $B_{\rm HP}$ following NV hyperpolarisation, compared to simply detecting $B_{\rm rms}$. In most NV experiments, the noise scales as $T_{\rm meas}^{-1/2}$ where $T_{\rm meas}$ is the time dedicated to the NV measurement which can be written as $T_{\rm meas}=\beta T$ where $T$ is the total experimental time and $\beta$ is the duty cycle of the measurement, which depends on the details of the experiment including readout time, dead times etc~\cite{Barry2020}. For the same total time $T$, the SNR ratio is then 
\begin{eqnarray} 
\frac{{\rm SNR}_{\rm HP}}{{\rm SNR}_{\rm rms}} &=&\frac{\Delta p_{\rm NV}^{\rm HP}}{\Delta p_{\rm NV}^{\rm rms}} \sqrt{\frac{\beta_{\rm HP}}{\beta_{\rm rms}}} \nonumber  \\ 
&\sim & \frac{\gamma_e B_{\rm HP}\tau_{\rm sens}^{\rm HP}}{(\gamma_e B_{\rm rms}\tau_{\rm sens}^{\rm rms})^2} \sqrt{\frac{\beta_{\rm HP}}{\beta_{\rm rms}}}  \label{eq:SNR}
\end{eqnarray}
which captures the ratio of the signal and the ratio of the noise based on the above discussion, with $\beta_{\rm HP}$ ($\beta_{\rm rms}$) the duty cycle for the hyperpolarised (statistically polarised) case. 
In Eq.~\ref{eq:SNR}, $\tau_{\rm sens}^{\rm HP,rms}$ is the optimum sensing time for each case, which we take to be $\tau_{\rm sens}=T_{\rm 2,NV}$ or the $\tau_{\rm sens}$ giving $\Delta p_{\rm NV}=1$, whichever is shortest. This ensures that the signal is not greater than it can be in reality. 
  
Because of the different exponents, the ratio of signals $\frac{\Delta p_{\rm NV}^{\rm HP}}{\Delta p_{\rm NV}^{\rm rms}}$ can easily exceed unity even when $B_{\rm HP}$ is comparable to or smaller than $B_{\rm rms}$. However, the ratio of the duty cycles is generally very unfavourable to the hyperpolarisation pathway. Indeed, in this case the experimental sequence adds a polarisation step (which takes $\sim T_{1,n}$) and a $\pi/2$ RF pulse (10's of $\mu$s, neglected in what follows) before each NV measurement. The duration of a single measurement is limited by the correlation time of the AC magnetic field, which is typically in the range $\tau_c\sim10-100~\mu$s for shallow NV centres ($d_{\rm NV}\sim5$~nm) both for solid and liquid samples~\cite{Staudacher2015}.     
This means that the measurement would take up only a fraction $\beta_{\rm HP}\sim\frac{\tau_c}{T_{1,n}}$ of the total time. Assuming ${\beta_{\rm rms}}$ is close to unity, we obtain
\begin{eqnarray} \label{eq:SNR2}
\frac{{\rm SNR}_{\rm HP}}{{\rm SNR}_{\rm rms}} &\sim & \frac{\gamma_e B_{\rm HP}\tau_{\rm sens}^{\rm HP}}{(\gamma_e B_{\rm rms}\tau_{\rm sens}^{\rm rms})^2} \sqrt{\frac{\tau_c}{T_{1,n}}}.
\end{eqnarray} 
Given that $B_{\rm HP}$ increases with increasing $T_{\rm 2,NV}$ and $T_{1,n}$, one can expect a maximum for $\frac{{\rm SNR}_{\rm HP}}{{\rm SNR}_{\rm rms}}$ as a function of these parameters. This is explored in Fig.~\ref{Fig_nanoNMR}b, which plots the SNR ratio as a function of $T_{1,n}$ for different densities $\rho_n$ and different NV coherence times $T_{\rm 2,NV}$, assuming a correlation time $\tau_c=100~\mu$s. The SNR ratio is always below 1, showing that the increased signal is not sufficient to overcome the increased noise. The most promising regime is that of small densities, e.g. $\rho_n=0.06$~nm$^{-3}$ i.e. 100 mM, and short coherence times, e.g. $T_{\rm 2,NV}=10~\mu$s, for which the magnetic fields are small hence the ratio of signals can be quite large, in the range 10-100. On the other hand, large densities and long $T_{\rm 2,NV}$ means that the signals are often saturated, $\Delta p_{\rm NV}^{\rm AC}\sim\Delta p_{\rm NV}^{\rm rms}\sim 1$, which leads to an SNR ratio $\ll 1$. 

We also examined the case of a sample confined in the vertical direction, instead of semi-infinite sample as assumed before. This could apply, for instance, to a lipid bilayer (a few nanometres thick)~\cite{Ishiwata2020} or an atomically-thin van der Waals material~\cite{Lovchinsky2017}. In this case, the polarisation extends further in the lateral directions (parallel to the diamond surface), however, we found that the magnetic field $B_{\rm HP}$ seen by the NV remains relatively unchanged. On the other hand, confinement in three dimensions can have a measurable effect. For example, a 10~nm~$\times$~10~nm~$\times$~5~nm sample positioned above the NV would generate a field $B_{\rm HP}$ larger by a factor $\approx10$ compared to a semi-infinite sample, assuming $^1$H spins with $\rho_n=66$~nm$^{-3}$, $d_{\rm NV}=5$~nm and $\theta_{\rm NV}=54.7^\circ$. Nevertheless, the SNR ratio $\frac{{\rm SNR}_{\rm HP}}{{\rm SNR}_{\rm rms}}$ remains below unity in all cases.



We note that the spectral resolution in nano-NMR experiments is also unlikely to be improved by NV hyperpolarisation. Indeed, the spectral resolution is limited by the correlation time $\tau_c$ of the oscillating/fluctuating magnetic field detected by the NV, which is expected to be similar whether it is the fluctuating field from statistical polarisation or the AC field from net polarisation. This resolution limit can be readily reached through correlation spectroscopy for statistical polarisation~\cite{Laraoui2013,Staudacher2015,Aslam2017}, and through FID-like measurements for net polarisation~\cite{Glenn2018}. For solid samples, $\tau_c$ is given by the dephasing time of the nuclear spins ($T_{2,n}^*$), which is governed by dipolar interactions. In conventional solid-state NMR, this dipolar broadening is often efficiently removed by magic-angle spinning and homo- or heteronuclear decoupling sequences, but in principle these methods can be applied to statistical polarisation as well~\cite{Aslam2017}.

\section{Conclusion} \label{sec:conclusion}

In this work, we theoretically investigated several potential applications of nuclear spin hyperpolarisation based on optically pumped NV centres in diamond. We first analysed the possibility of using NV hyperpolarisation to polarise a macroscopic sample that would then be measured in a conventional NMR spectrometer. We found that, for NV hyperpolarisation to be competitive with existing hyperpolarisation techniques, a key condition is to specially engineer the diamond structure to maximise the contact area between NVs and sample.
We predicted, assuming optimally efficient polarisation transfer, that enhancements over thermal polarisation by up to two orders of magnitude can be obtained with existing technology. Larger enhancements, equivalent to polarisations exceeding 10\%, can in principle be obtained but this requires structuring the diamond with aspect ratios of over a hundred, which is an outstanding challenge. We discussed factors reducing the polarisation transfer efficiency, especially the finite NV spin coherence time, which motivates further work in the optimisation of diamond materials. We also outlined some of the practical challenges of NV hyperpolarisation, such as the need for high power laser illumination, and the requirement that the sample be in a solid form for the hyperpolarisation step. Overall, this application emerges as challenging, but the prospect of realising a versatile, non-invasive hyperpolarisation platform at a fraction of the cost of existing techniques warrants further work.

Next, we examined the possibility of integrating NV hyperpolarisation into NV-based liquid-state micro-NMR. NV-based micro-NMR is a recently developed technique~\cite{Glenn2018} that could lead to the realisation of portable NMR spectrometers. The technique relies on NV centres located several $\mu$m away from the diamond surface, which limits the NMR sensitivity. We found that, by adding a layer of near-surface NV centres to hyperpolarise the sample, a signal boost of about 1-2 orders of magnitude over thermal polarisation can be obtained in principle, with further enhancements expected with micro-structuring of the diamond surface. Thus, built-in NV hyperpolarisation could prove a convenient way to boost the sensitivity of NV micro-NMR, without the inconvenience of conventional DNP which requires adding free radicals to the solution~\cite{Bucher2020}. One drawback of NV hyperpolarisation, however, is the need to freeze the sample for the hyperpolarisation step, which adds a technical complication and may not be desirable/possible for some samples. It should also be noted that our predictions are based on a simplistic model where the polarisation distribution is assumed to be unchanged upon thawing of the sample. Further work will be required to test more sophisticated models taking into account polarisation diffusion during the NMR measurement in the liquid state.

Finally, we analysed the case of nano-NMR, where near-surface NV centres are normally employed to detect the randomly fluctuating magnetic field induced by statistical polarisation. We found that using the same NV to hyperpolarise the sample and measure the net polarisation instead can lead to an increase in the measurable signal. However, this generally does not translate into a net improvement in the signal-to-noise ratio of the measurement because of the significant temporal overhead of the hyperpolarisation step, which dominates the comparatively short measurement time typically involved in nano-NMR experiments. Thus, NV hyperpolarisation for nano-NMR seems to be the least promising application, although future work could look at techniques that may mitigate the impact of the temporal overhead, such as quantum-memory-assisted repetitive readout schemes~\cite{Lovchinsky2016,Hopper2018}.

Another area of interest for future work is the possibility to make the polarisation transfer more efficient by exploiting a different pathway to the direct NV-sample coupling studied here. For instance, Ref.~\cite{Zangara2019b} theoretically investigated the possibility of exploiting ancillary electron spins to enhance the coupling, which in effect amounts to reducing the NV-target distance. Another intriguing idea is to use the natural bath of $^{13}$C spins present inside the diamond as a polarisation buffer~\cite{Ajoy2018}. In this scheme, a single NV spin would polarise a large number of those internal $^{13}$C spins over the course of minutes to hours (limited by the longitudinal relaxation time of $^{13}$C in diamond), and this internal $^{13}$C polarisation would then be transferred to (or spontaneously diffuse towards) the target nuclear spins located outside the diamond. Assuming the target spins have a much shorter $T_{1,n}$ time than the internal $^{13}$C, this amounts to increasing the density of polarising agents in the diamond from the NV density ($\sim10$~ppm) to the internal $^{13}$C density (1.1\% for natural isotopic concentration), i.e. a 100-fold increase. However, further work is required to quantify the potential of this approach in realistic conditions.

\section*{Acknowledgements}

We acknowledge support from the Australian Research Council (ARC) through grants DE170100129, CE170100012 and DP190101506. A.J.H. is supported by an Australian Government Research Training Program Scholarship. 

\bibliographystyle{naturemag}
\bibliography{bib}

\clearpage

\end{document}